\documentclass[useAMS,usenatbib,a4paper]{mn2e}
\usepackage{xcolor}
\usepackage{graphicx, amsmath, amssymb}

\title[Short GRBs with EE Observed with BAT and GBM]{Short Gamma-Ray Bursts with Extended Emission Observed with
{\it Swift}/BAT and {\it Fermi}/GBM}

\author[Y. Kaneko, Z.F. Bostanc\i, E. G\"o\u{g}\"u\c{s}, L. Lin]
{Y. Kaneko,$^{1}$ Z.F. Bostanc\i,$^{1,2}$ E. G\"o\u{g}\"u\c{s},$^{1}$ L. Lin$^{1,3}$ \\
  $^{1}$Sabanc\i~University, Faculty of Engineering and Natural
  Sciences, Orhanl\i-Tuzla 34956 Istanbul Turkey\\
  $^{2}$Istanbul University, Faculty of Sciences, Department of
  Astronomy and Space Sciences, University 34119 Istanbul, Turkey \\
  $^{3}$APC laboratory-FACe, 13 Rue Watt 75205 Paris Cedex 13 France}
  \begin{document}
  \date{}
  \maketitle
  \label{firstpage}

  \begin{abstract}
Some short GRBs are followed by longer extended emission, lasting
anywhere from $\sim$10 to $\sim$100 s.   These short GRBs with
extended emission (EE) can possess observational characteristics of
both short and long GRBs (as represented by GRB 060614), and the
traditional classification based on the observed duration places
some of them in the long GRB class.  While GRBs with EE pose a
challenge to the compact binary merger scenario, they may therefore
provide an important link between short and long duration events. To
identify the population of GRBs with EE regardless of their initial
classifications, we performed a systematic search of short GRBs with
EE using all available data (up to February 2013) of both {\it
Swift}/BAT and {\it Fermi}/GBM.  The search identified 16 BAT and 14
GBM detected GRBs with EE, several of which are common events
observed with both detectors.  We investigated their spectral and
temporal properties for both the spikes and the EE, and examined
correlations among these parameters. Here we present the results of
the systematic search as well as the properties of the identified
events. Finally, their properties are also compared with short GRBs
with EE observed with BATSE, identified through our previous search
effort.  We found several strong correlations among parameters,
especially when all of the samples were combined. Based on our
results, a possible progenitor scenario of two-component jet is
discussed.
  \end{abstract}


\section{Introduction}
The progenitors of extremely energetic gamma-ray bursts (GRBs) are
most likely either the core collapse of a massive star
\citep[collapsar;][]{woosley1993,mac01} or the coalescence of two
compact objects (neutron stars or black holes) in a binary system
\citep[see e.g.,][]{eichler89}. An increasing number of
observational facts point toward association of long GRBs (longer
than a few seconds) with massive stars collapsing into black holes
\citep{wb06} or possibly fast-rotating magnetars \citep{tho04}. In
particular, observational predictions of the collapsar scenario
include association with core-collapse supernovae (SNe), and also
with late-type host galaxies having active star formation.

On the other hand, for short GRBs (lasting less than a few seconds),
observations so far have yielded no support, even contradictions with the
collapsar scenario. There has been no SN association with any short
GRBs albeit some extensive search efforts were in place
\citep{ber14}. Short GRBs seem to occur relatively nearby (with cosmological redshift, $z
\lesssim$ 1), and their host galaxies can be both early-type and
late-type.  Moreover, the short emission timescales and the
energetics involved pose challenges for the collapsar model.
Therefore, the compact-binary merger scenario remains to be the most
plausible picture to account for their observational properties.

For intermediate-duration GRBs ($\sim$5 s), however, the
classification may not be clear-cut: over the last decade, a growing
population of short GRBs with extended emission (EE) has emerged.
These events are characterized with an initial short burst phase,
followed by a much longer ($\gtrsim$ few tens of seconds) emission
episode. EE components are usually (but not always) relatively
dimmer and spectrally softer compared to the initial short spike.
GRBs with EE exhibit several spectral and temporal properties of
both short and long GRBs, therefore, could play a pivotal role in
understanding the nature of both canonical burst types.

Since the first clear identification of such ``hybrid" GRBs in 2006
\citep[GRB\,060614;][]{gehrels06}, there has been a few extensive
investigations to search for more of these type of events and
uncover their observational characteristics. Norris \& Bonnell
(2006) have identified $\sim$15 GRBs with EE via visual inspection
of $\sim$2700 bursts observed with Burst and Transient Source Experiment (BATSE)
on board {\it Compton Gamma-Ray Observatory}, among which 8
GRBs were investigated. They found that the spectral lags for the
initial spikes of these events were negligible. Bright long GRBs, on
the contrary, usually display noticeable lags, with an average of
$\sim$50$-$100\,ms \citep{norris00}.  All of these events were
classified as long GRBs with the durations, $T_{90}>$ 2\,s in the
BATSE catalog.
In addition, extended emission was also identified in a dozen
short-duration GRBs observed with {\it Swift}/Burst Alert Telescope
(BAT), in the Bayesian block representation of their lightcurves
\citep{norris10,norris11,sak11}. The initial spikes of these events
are also spectrally harder and displayed no spectral lags. Most of
them were associated with X-ray afterglow observed with {\it
Swift}/X-Ray Telescope (XRT), that are brighter and longer lasting
than the afterglow of non-EE GRBs.

We previously searched systematically the entire BATSE GRB dataset to
 identify GRBs with EE that are similar to GRB\,060614
\citep{bos13}.  We found 18 such GRBs with clear EE components
within the BATSE data, and investigated their properties. Here, we
extend our search to include all GRBs observed with {\it
Fermi}/Gamma-ray Bust Monitor (GBM) and {\it Swift}/BAT before
February 2013 (regardless of their $T_{90}$ duration).
A few notable advantages of extending the search to GBM and
BAT events are:
\begin{enumerate}
\item It increases the total sample size of GRBs with EE, providing
better statistics for examining their overall properties,
\item About 80\% of all BAT GRBs have afterglow observation with XRT, and 30\% have measured redshift values; if
found, the afterglow and the source-frame properties of GRBs with EE
can be investigated, and
\item There are many GRBs observed simultaneously with GBM and BAT; if such common GRBs are associated with EE, more in-depth information about the EE could be obtained, as well as shed light on systematic
differences between the instruments.
\end{enumerate}

In this paper, we present the results of our systematic search for
short GRBs with EE using the data of {\it Swift}/BAT and {\it
Fermi}/GBM, and subsequent analysis of the events identified with
the search.  Below, the event selection and the search methodology
are described in \S\ref{sec:method}, and the search results and
investigation of the detection sensitivity are presented in
\S\ref{sec:results}.  Then in \S\ref{sec:analysis}, we present the
properties of the candidate events based on our analysis. Finally,
we discuss the implication of our results in \S\ref{sec:discussion}.

\section{Event Selection and Search Methodology}\label{sec:method}
We first selected all events that consist of initial short spikes,
and the selected events were then subjected to the EE search described in detail below.  We
used all available data up to January 31, 2013 for both BAT and GBM.
We applied essentially identical search methodology to both BAT data
and GBM data, which was also used for our previous systematic search
with BATSE data \citep{bos13}. However, the data coverage, reduction
procedures, and the data types are different for the two
instruments, and therefore, slight adjustments were made according
to the detector characteristics, which we describe in this section.

\subsection{Data and Event Selection}
For both BAT and GBM events, the first sample selection was made
based on the burst duration ($T_{90}$) published in their catalogs
\citep{pac12,sak11}. We note that $T_{90}$ is calculated in slightly
different energy bands for the two detectors; 15$-$350\,keV for BAT
and 50$-$300\,keV for GBM, due to the difference in the
sensitivities and energy coverage of the detectors. All short GRBs
with $T_{90}\le$ 5\,s were included in the systematic search for EE.
Additionally, to identify and include in the search all GRBs with
correct morphology (i.e., containing initial short spikes regardless
of their $T_{90}$), we first subjected all of the long GRBs to a
morphological test as described next. \\

\noindent{\bf BAT:} There are 685 GRBs detected with BAT between 20
November 2004 and 31 January 2013, among which 95 are short
($T_{90}\le$ 5\,s) and 590 are long GRBs ($T_{90} >$ 5\,s).

For all of the long events, we extracted background-subtracted
(i.e., mask-weighted) lightcurves in 15$-$350\,keV with 64-ms time
resolution, using {\tt batbinevt} included in HEASOFT version 6.12.
The count rates were then multiplied by the number of enabled
detectors for each event to obtain the total count rates.

\vspace{6pt} \noindent {\bf GBM:} There are 1040 GRBs detected with
GBM between 1 August 2008 and 31 January 2013, among which 246 are
short ($T_{90} \le$ 5\,s) 794 are long GRBs ($T_{90} >$ 5\,s).

For each of the long GRBs, we used Time-Tagged Event (TTE) burst
data binned to 64-ms time resolution, and subtracted the background
rate, which is an average count rate of a pre-burst interval ($\sim
T_0$$-$30\,s to $T_0$$-$2\,s, where $T_0$ is the burst trigger time).\\

For both BAT and GBM, we then subjected the lightcurves of the long
GRBs to the following morphological criteria:
\begin{enumerate}
\item The burst peak occurs before $T_0$+5\,s, and
\item The count rates remain below 30\% (or 40\% for some BAT events with low
peak rate of $<$11k count/s) of the peak count rate for at least
50\% of the rest of the duration after the peak time until
$T_0$+5\,s.
\end{enumerate}

 We found 33 BAT
(103 GBM) events out of 590 (808) long GRBs that matched the
morphological criteria. We then searched for EE in these events
together with the 95 (184) short GRBs.  The total numbers of events
subjected to the search was 128 for BAT and 287 for GBM.

\subsection{Background Determination}
Correct background modeling is crucial in detecting the EE
components, as EE can be weak and only slightly above the
background. For BAT, the background is not an issue since the
lightcurve produced is mask-weighted, meaning that the lightcurve is
essentially background free if there is no other bright source (usually known)
in the field of view.

\begin{figure}
\centering
\includegraphics[scale=0.42]{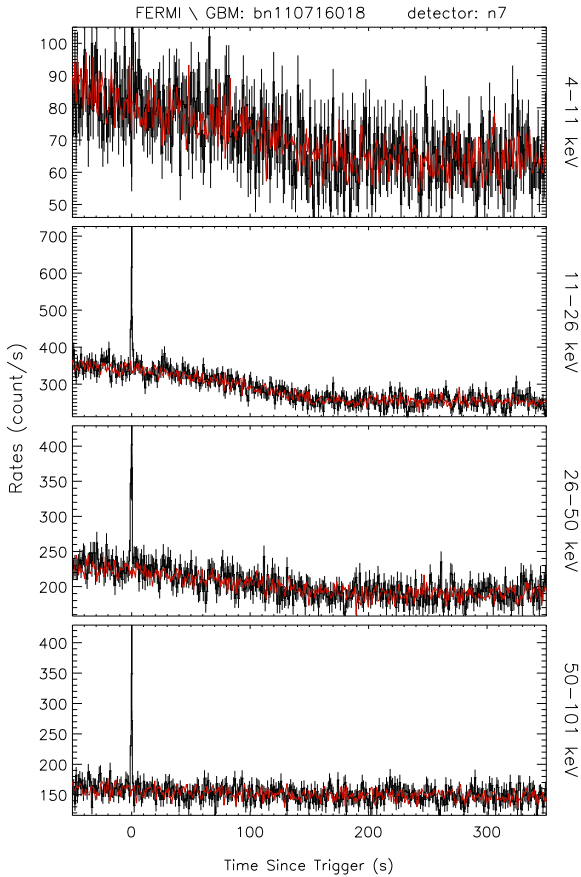}
\caption{An example GBM lightcurve in four energy bands (indicated on the right
side of each panel), showing the orbital background (red).} \label{background}
\end{figure}

For GBM, on the other hand, the background rates at the time of a
transient event have to be estimated by modeling, based on its
long-term behavior in all energy band.  The GBM background can be
quite variable over a relatively-long period of time, which could
hinder a simple polynomial interpolation. Therefore, as suggested by
\citet{fit11} and as used also for our previous EE search with the
BATSE data, we determined background rates for GBM events from the
data of adjacent days. The {\it Fermi} satellite was at the same
coordinates every $\sim$15 orbits (corresponding to $\sim$24 hours),
and the spacecraft rocking angle was the same every 2 orbits.  Thus,
we used either the average of $T_0 \pm 30$ orbits or the average of
$T_0 \pm 14$ and $T_0 \pm 16$ orbits \citep[see also][]{fit11}.
 When both types of orbital data were available,
we compared the average rms of the count rates over a pre-burst
interval of $T_0$$-$50 to $T_0$$-$10 s of the triggered and the
orbital data, and the orbital background that better matched the
data rms was used for the search. An example of such orbital
background lightcurve is shown in Figure~\ref{background}.

Using the orbital background, we generated the background-subtracted
lightcurves for 257 GRBs out of the 287 GBM GRBs.  There were 30
GRBs for which the data of the previous orbits were either not
available or incomplete, and the orbital background could not be
used.  Visual inspection of these events revealed two bursts that
clearly showed extended emission, so we used low-order polynomial
interpolation to model the background rates for the two events,
which are indicated accordingly in Table~\ref{events}.  The other 28
bursts were excluded from the sample.
Additionally, for one event also showing EE (GRB\,121029), the
orbital background was available but clearly not matching;
therefore, the background was modeled with a polynomial function
around the burst time also for this event.

\subsection{EE Search Criteria and Method}
Our EE search is based on Signal-to-Noise Ratio (SNR) of binned
lightcurve. The EE search criteria we employed here were identical
to those used in our previous search with BATSE data. The criteria
were defined based on previously-observed short GRBs with EE, such
as GRB\,060614 \citep{gehrels06} and those found by
\citet{norris06}, all of which seem to constitute the hybrid class
of short and long GRBs. We first extracted energy-resolved
lightcurves for all of the GRBs in our sample, as follows:

\vspace{6pt} \noindent {\bf BAT:} We extracted 1-s resolution
lightcurves both background-subtracted and with background (i.e.,
unweighted), for each of the 128 GRBs in four energy bands; 15$-$25,
25$-$50, 50$-$100, and 100$-$150\,keV. We used the unweighted
lightcurves for determining the background fluctuation ($\sigma$
values).

\vspace{6pt} \noindent {\bf GBM:} For each of the 260 GRBs, we
subtracted background for two detectors in which the event appeared brightest, using daily CTIME
data rebinned to 1-s resolution.  The CTIME data provide 8
energy channels in $\sim$5$-$2000\,keV. \\

For both BAT and GBM, we subsequently binned the lightcurves with
4-s resolution between $T_0+5$ and $T_0+350$\,s and calculated the
SNR of each bin. We used the lightcurves of two energy bands:
$<$50\,keV and 50$-$100\,keV.  The lowest energy thresholds were 15
for BAT, and $\sim$10\,keV for GBM. The search algorithm positively
identified EE when SNR $\ge 1.5\sigma$ above background for at least
consecutive 12\,s (3 bins) in any of these energy bands. In case of
GBM, the criteria had to be met in both detectors for the positive
identification of the EE, as GBM data did not provide specific
directional information. We further checked the lightcurves of all
12 NaI detectors of GBM candidates so as to exclude the possibility
that the detected EE was due to another soft transient source.

\section{Search Results and Detection Sensitivities}\label{sec:results}
 Using these criteria, we found 37 BAT GRBs and
72 GBM GRBs as candidates for short GRBs with EE. We subsequently
inspected each one of these candidates, and manually eliminated ones
with obvious background issues (e.g., Earth occultation of another
source or data mismatch over longer period), as well as those that
are clearly multi-episodic long GRBs based on the morphology and the
spectral hardness.  We additionally excluded the events whose EE was
not visible at all in any other energy range, to reduce the
possibility of chance detection.

After the manual filtering, we were left with 16 BAT GRBs and 14 GBM
GRBs as the true EE candidates. We list all of the identified events
in Table~\ref{events} with their $T_{90}$, durations of the spikes
and the EE (see \S\ref{sec:duration}), and afterglow observation
information where available. The lightcurves of all of these
candidate events are available online,\footnotemark{} one of which
is shown as an example in Figure~\ref{bb_lc}. As seen in
Table~\ref{events}, most of the BAT events with EE have associated
afterglow observations in x-ray and optical bands, and 8 events have
their redshift values available. There are 3 events for which X-ray
flares are seen in their X-ray lightcurves observed with {\it Swift}
X-Ray Telescope (XRT).  One event's flare actually coincides with
the EE detected here (GRB\,100212A; see \S\ref{sec:spec} for the
discussion).  The redshift values are mostly $\lesssim$1, which is
consistent with short GRBs' redshift distribution \citep{ber14}. One
exception to this is GRB070506 with $z=2.31$.  This is an
exceptionally high $z$ value for a short GRB.
\footnotetext{Available at \\
http://people.sabanciuniv.edu/yuki/GRBEE\_BAT\_GBM\_all.pdf}

Among these, 5 GRBs were common events observed both with BAT and
with GBM, which we also indicate in Table~\ref{events} with C1 or
C2. For two of the five common events, EE was only detected in BAT
in our search. The common events were additionally subjected to
joint spectral analysis and discussed separately in
\S\ref{sec:common}.  Our search also detected EE
components in the GBM data of two additional common events,
GRB\,090518 and GRB\,091127, extending up to $\sim$300\,s and
$\sim$15\,s after the trigger, respectively. However, we found out
that GRB\,090518 EE was an activity from Vela X-1 coinciding with
the GRB direction (C.\,Wilson-Hodge, private communication), which
was flagged by the fact that the BAT lightcurve showed no indication
of the EE. GRB\,091127 was a long GRB associated with SN\,2009nz
\citep{cob10}.
Thus these events were excluded from the list of candidates.

We note that although these 30 candidate GRBs presented in
Table~\ref{events} all cleared the search criteria, some of them may
be of different nature based solely on their lightcurves. They all
contain initial short spikes of $\lesssim$5\,s indeed; however,
there are cases where, for example, there is a preceding weaker peak
that makes the actual duration of the spike longer (GRB\,090131), or
the entire burst is pretty spiky that the subsequent ``extended
emission" appears to be rather a continued episode of a long GRB
(GRB\,091120). Nonetheless, as it is difficult to classify events
only from their lightcurves, we did not exclude these events from
the list.
In addition, for one of the BAT events (GRB\,051016B), Vela X-1 with varying flux was in the BAT field-of-view at the time of the burst trigger.  Therefore, we caution that the possibility of contribution of this X-ray source to the EE detected in GRB\,051016B cannot be ruled out.
All of these candidate events were subjected to the
subsequent analysis.

Finally, the candidates also include 7 of the 12 BAT GRBs with EE
previously identified by \citet{norris10} and \citet{sak11}. The
other 5 were not found in our search because the extended emission
components of 4 of them were too short and too weak for our search
criteria, and one of them had a spike $>$5\,s so did not meet our
morphological criteria. We note that albeit short and weak, there
are visible indications of extended emission in the lightcurves of
these 5 events.
\begin{table*}
\centering \setlength{\tabcolsep}{0.07in} \caption{Properties of the
candidates for GRBs with EE identified in the search.  The duration
of the spike and the EE were determined using Bayesian block method.
 C1 and C2 indicate the common events observed with both BAT and GBM.
 C1 events' EE were only identified in either BAT or GBM data whereas
 C2 events' EE were identified in both datasets.}
\begin{tabular}{rlrrrrrrccc}
\hline
&GRB name & ${T_{0}}$ Time   & ${T_{90}}^a$  & $T_{\rm spike}$ & $T_{\rm EE}$ & $B_{\rm spike}$$^{b}$& $B_{\rm EE}$$^{b}$ & Afterglow$^{c}$ & $z$ &   X-ray flare  \\
                 &               &  UT     &   (s) &      (s) &  (s) & (s)    &  (s)     &      &    &  \\
\hline
&{\bf BAT}&      &       &       &       &       &       &       &     &    \\
    &   050724$^d$    &   12:34:09    &   96  &   2.76    &   107 &   $-$0.02   &   3.04    &   XOR &   0.258   &   $T_0$+10$^4s$   \\
    &   051016B &   18:28:09    &   4   &   4.03    &   33  &   0.07   &   4.23   &   XO  &   0.9364  &   $T_0$+409s  \\
    &   060614$^d$    &   12:43:49    &   108.7   &   5.89    &   169 &   -1.55   &   7.24    &   XO  &   0.125   &   --  \\
    &   061006$^d$    &   16:45:51    &   129.9   &   2.05    &   113 &   -23.2 &   2   &   XO  &   0.4377  &   --  \\
    &   061210$^d$    &   12:20:39    &   85.3    &   0.13    &   77  &   0.21    &   1.04    &   X   &   0.4095  &   --  \\
    &   070506  &   5:35:58 &   4.3 &   5.25    &   15  &   3.75   &   38  &   XO  &   2.31    &   --  \\
    &   070714B$^d$   &   4:59:29 &   64  &   2.88    &   39  &   $-$0.8    &   32.29   &   XO  &   0.92    &   --  \\
    &   080503$^d$    &   12:26:13    &   170 &   0.38    &   147 &   0.11   &   6   &   XO  &   --  &   --  \\
C1  &   090531B$^d$   &   18:35:56    &   80  &   1.02    &   54  &   0.29    &   2.04    &   X?  &   --  &   --  \\
    &   090927& 10:07:16    &   2.2 &   2.18    &   28  &   0.06   &   2.95    &   XO  &   1.37    &   --  \\
C1  &   100212A$^{e}$& 14:07:22    &   136 &   2.18    &   135 &   $-$0.32   &   1.86    &   XO  &   --  &   $T_0$+$\sim$100s    \\
C2  &   100522A &   3:45:52 &   35.3    &   3.97    &   15  &   $-$0.68   &   23.86   &   X   &   --  &   --  \\
C2  &   110207A &   11:17:20    &   80.3    &   3.07    &   137 &   0.184   &   10.6    &   --  &   --  &   --  \\
C2  &   110402A &   0:12:57 &   60.9    &   2.52    &   82  &   3.56    &   6.08    &   XO  &   --  &   --  \\
    &   111121A &   16:26:24    &   119 &   2.37    &   61  &   0.04    &   4   &   X   &   --  &   --  \\
    &   121014A &   20:11:56    &   80  &   0.96    &   81  &   $-$0.10&  14.8    &   --  &   --  &   --  \\
        &{\bf GBM}      &                     &              &              &             &              &             &             &              &   \\
    &   080807$^f$&23:50:33 &   19.07   &   1.28    &   27  &   $-$0.12   &   1.15    &   --  &   --  &   --  \\
    &   090131  &   2:09:21 &   35.07   &   7.81    &   23  &   1.98    &   22.27   &   --  &   --  &   --  \\
    &   090820  &   12:13:17    &   15.3            &     0.77  &   7   &   -0.19   &   8.32    &   --  &   --  &   --  \\
    &   090831  &   07:36:37    &   39.42   &   0.45    &   86  &   $-$0.12   &   0.32    &   --  &   --  &   --  \\
    &   091120  &   4:34:40 &   50.18   &   2.82    &   52  &   0.00    &   2.81    &   --  &   --  &   --  \\
    &   100517  &   3:42:08 &   30.46   &   1.41    &   11  &   $-$0.12   &   22.02   &   --  &   --  &   --  \\
C2  &   100522  &   3:45:52 &   35.33   &   3.97    &   13  &   $-$0.64   &   26.49   &   X   &   --  &   --  \\
C2  &   110207  &   11:17:20    &   37.89   &   3.07$^{g}$ &   38  &   $-$0.32   &   2.68    &   --  &   --  &   --  \\
C2  &   110402  &   0:12:58 &   35.65   &   1.41$^{g}$ &   39  &   $-$0.32   &   1.08    &   XO  &   --  &   --  \\
    &   110824  &   0:13:10 &   76.61   &   1.54    &   93  &   0.00    &   1.54   &   --  &   --  &   --  \\
    &   111228  &   10:52:50    &   2.94    &   6.53    &   30  &   $-$0.51   &   25.7    &   --  &   --  &   --  \\
    &   120402  &   16:04:01    &   20.22   &   5.63    &   19  &  $-$2.30  &   4.16    &   --  &   --  &   --  \\
    &   120605$^f$& 10:52:16    &   18.11   &   2.82    &   8   &   $-$0.57   &   12.41   &   --  &   --  &   --  \\
    &   121029$^f$  &   8:24:20 &   15.81   &   0.38    &   6   &   $-$0.12   &   10.75   &   --  &   --  &   --  \\
\hline
\end{tabular}
\label{events}
 \footnotesize{ \begin{flushleft}
$^a$ Taken from \citet{sak11} and \citet{pac12} \\
$^{b}$ The begining of the spike and the EE since trigger time\\
$^{c}$ X = X-ray, O = optical, R = radio \\
The afterglow and redshift information was obtained from
J.\,Greiner's web page at http://www.mpe.mpg.de/\~jcg/grbgen.html as
well as from the XRT
lightcurve repository at http://www.swift.ac.uk/xrt\_curves\\
$^d$ Also identified as GRBs with EE in \citet{norris10} using
Bayesian block method \\
$^{e}$ Double-episodic event; see \S\ref{sec:spec} for more information \\
$^f$ Background rates were modelled with low-order polynomial interpolation\\
$^{g}$ Calculated using 1.024 resolution\\
\end{flushleft}}
\end{table*}


%
\subsection{Durations of Spikes and EE}\label{sec:duration}
It is noteworthy that none of our 30 candidate events can be
classified as ``short GRBs" by the conventional definition of
$T_{90}<2$\,s.  However, they all consist of short spikes, so we
systematically determined the durations of the spike and the EE
separately for each event with Bayesian block (BB) method
\citep{sca13}.  The BB representation simplifies the lightcurve with
a series of step function that could easily reveal local structures
in the time series. The duration, then, was the beginning and the
end of the spike or EE blocks.  We adopted the algorithm of
\citet{sca13} to generate a BB lightcurve for each event in the
energy range in which the EE was detected (usually $< 50$\,keV). We
used the mask-tagged data for BAT and background-subtracted
 CTIME data for GBM, with 64-ms and 1-s resolutions for this purpose.

\begin{figure}
\centering
\includegraphics[scale=0.4]{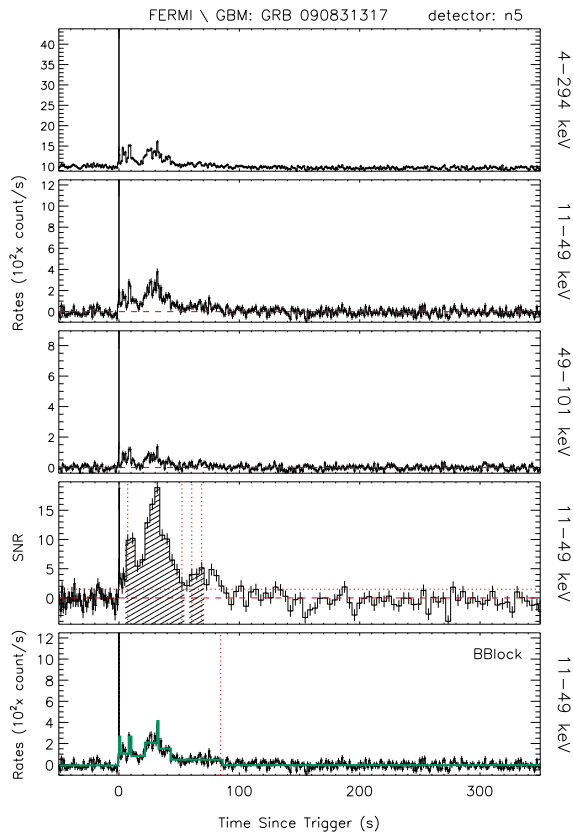}
\caption{An example GBM lightcurve in three energy bands as
indicated on the right side of each panel.  The bottom
  two panels are the binned SNR history and the Bayesian block
  lightcurve (green histogram), in which the EE selection is indicated
  in red vertical dashed lines.
  Further figures can be viewed in the electronic version of the article.$^1$} \label{bb_lc}
\end{figure}

  We first applied the BB method using the 1-s resolution data
   on the interval, 50~s before and 400~s after the burst trigger time
   for each burst to determine $T_{\rm EE}$.
   For the peak duration, we utilized the 64-ms data for both BAT and GBM,
   for the interval 40~s
  prior and 20~s after the burst trigger time. We could not
  measure peak duration of two GBM events (GRBs 110207 and 110402) due
  to insufficient statistic. Therefore, we listed peak durations of these
  two in 1-s domain in Table~\ref{events}. In cases
    where the end of the spike and the beginning of the EE were not
    clear-cut, we took the beginning of
    the block with the lowest intensity around $T_0$+5~s as the beginning of
    $T_{\rm EE}$.  In the lightcurves available from the links
given above, the BB lightcurve of each event is also shown with the
intervals of spike and EE indicated. An example is shown in
Figure~\ref{bb_lc}.

\begin{figure}
\centering
\includegraphics[trim=1cm 0cm 0cm 0cm, clip=true, scale=0.45]{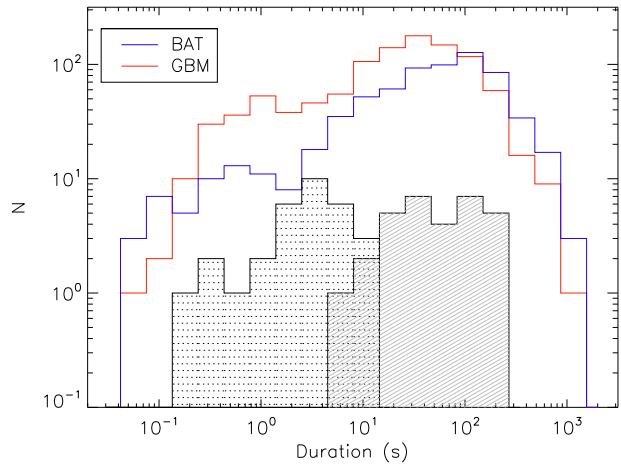}
\caption{Distribution of the duration of the spikes (dotted) and EE
(hashed) of all 30 candidates, shown against the $T_{90}$
distribution of all GRBs observed up to January 31, 2013 (BAT in
blue and GBM in red).} \label{duration}
\end{figure}

The resulting BB duration of the spikes and EE are presented in
Table~\ref{events} as $T_{\rm spike}$ and $T_{\rm EE}$ respectively,
along with the start time of the spike and the EE ($B_{\rm spike}$
and $B_{\rm EE}$). In Figure~\ref{duration}, we compare the
durations of the spikes and the EE components of our candidate
events to the $T_{90}$ distributions of all BAT and GBM GRBs
observed up to January 31, 2013.  The majority of our candidate
events have spikes that are longer than the 2-s division line, up to
$\sim$10\,s.  We caution, however, that the population comparison in
Figure~\ref{duration} is associated with a caveat: the GBM and BAT
durations are $T_{90}$ values (i.e., time during which 90\% of
photons are emitted) determined in the most sensitive energy range
of each instrument, in photon space for GBM \citep{pac12} and mostly
with the BB method for BAT \citep{sak11}. Here, our duration is the
total emission time in the energy range in which the EE was
detected.
 We further note that, the duration values of the 3 common GRBs with EE
 identified in both BAT and GBM data (GRBs\,100522A, 110207A, and 110402A),
 are not necessarily consistent between the BAT data and GBM
data.  This is because the EE components may have been identified in
different energy ranges, in addition to the difference between the
sensitivities of the two instruments.  So in the histogram, they are
included with their BAT duration and GBM duration independently.

\subsection{ Detection Rates and Sensitivities}\label{sec:sim}
To probe the sensitivity limits of both BAT and GBM and the rate of
the false-positive EE detection, we have simulated large sets of EE
lightcurves (LCs) and subjected them to our EE search algorithm,
which we describe in details below.

For each event and for each instrument, we first simulated the EE
using the actual Bayesian block representation of the EE LC (in the
energy band where the EE was identified), and added random noise to
the simulated EE.  The noise distribution used here is Gaussian for
BAT data and Poisson for GBM data, decided based on the real error
distributions. We then added the simulated, noise-added EE to the
``burst-free" background LC of the same energy range, which are:

\vspace{6pt} \noindent {\bf BAT:} mask-weighted LC of the time
interval from $T_0$+150 to 900\,s, displaced by $-$150 s (so the
background LC starts at $T_0$) for most of the bursts.  In the cases
of bursts with long EE or data with shorter time coverage, the time
interval was from $T_0+(1.2 \times T_{\rm EE,end}$) to MAX(data time).

\vspace{6pt} \noindent {\bf GBM:} real data for pre-/post- time
intervals which is the start time of the data to $T_0$--10 s and
$T_{\rm EE,end}$+10 s to the end time of the data. In between, we
use the background model fitted to the pre-/post-burst time
intervals using a 2nd order polynomial function.\\

\begin{figure}
\centering
\includegraphics[scale=0.4]{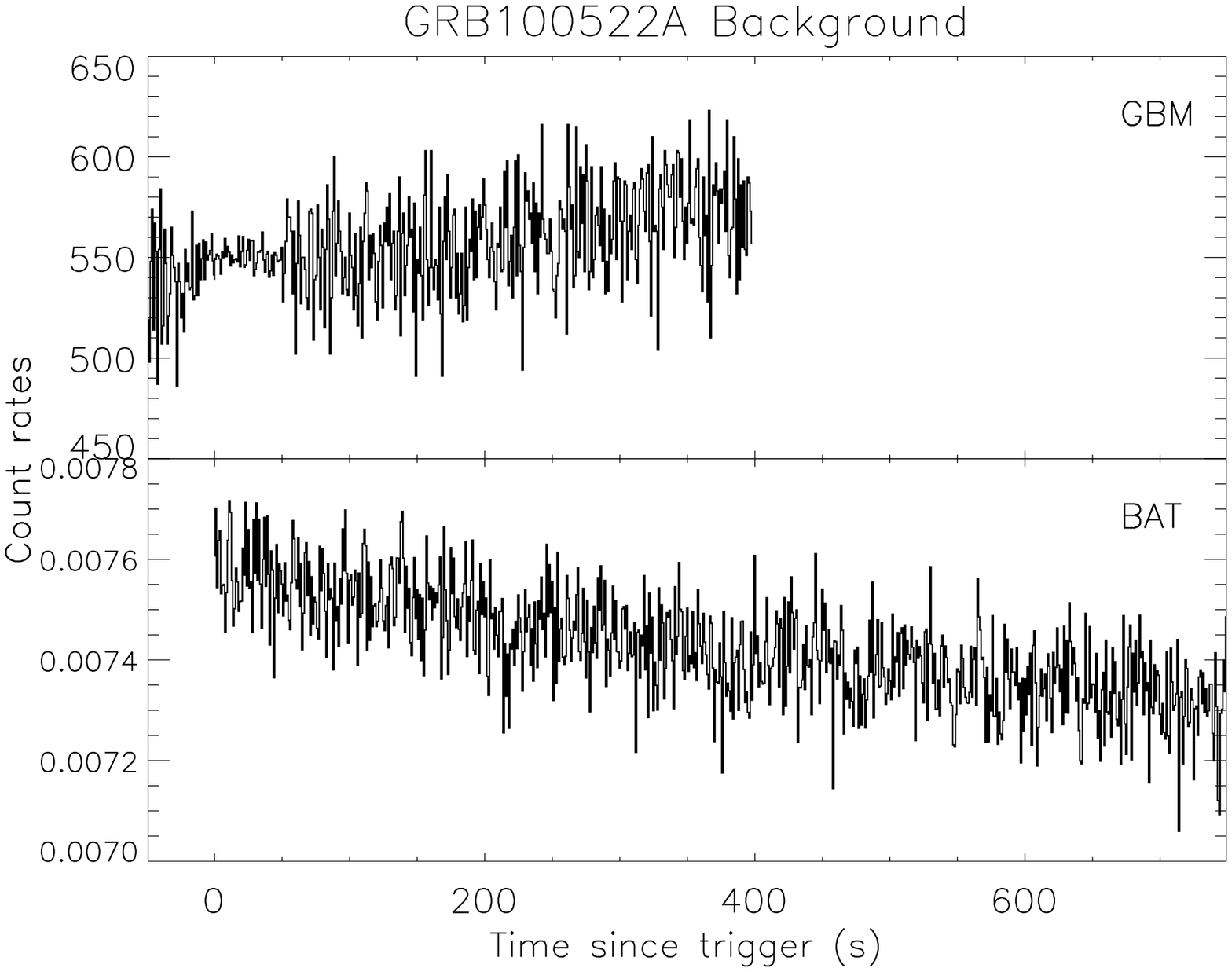}
\includegraphics[trim=0cm 0cm 0cm 2.6cm, clip=true, scale=0.4]{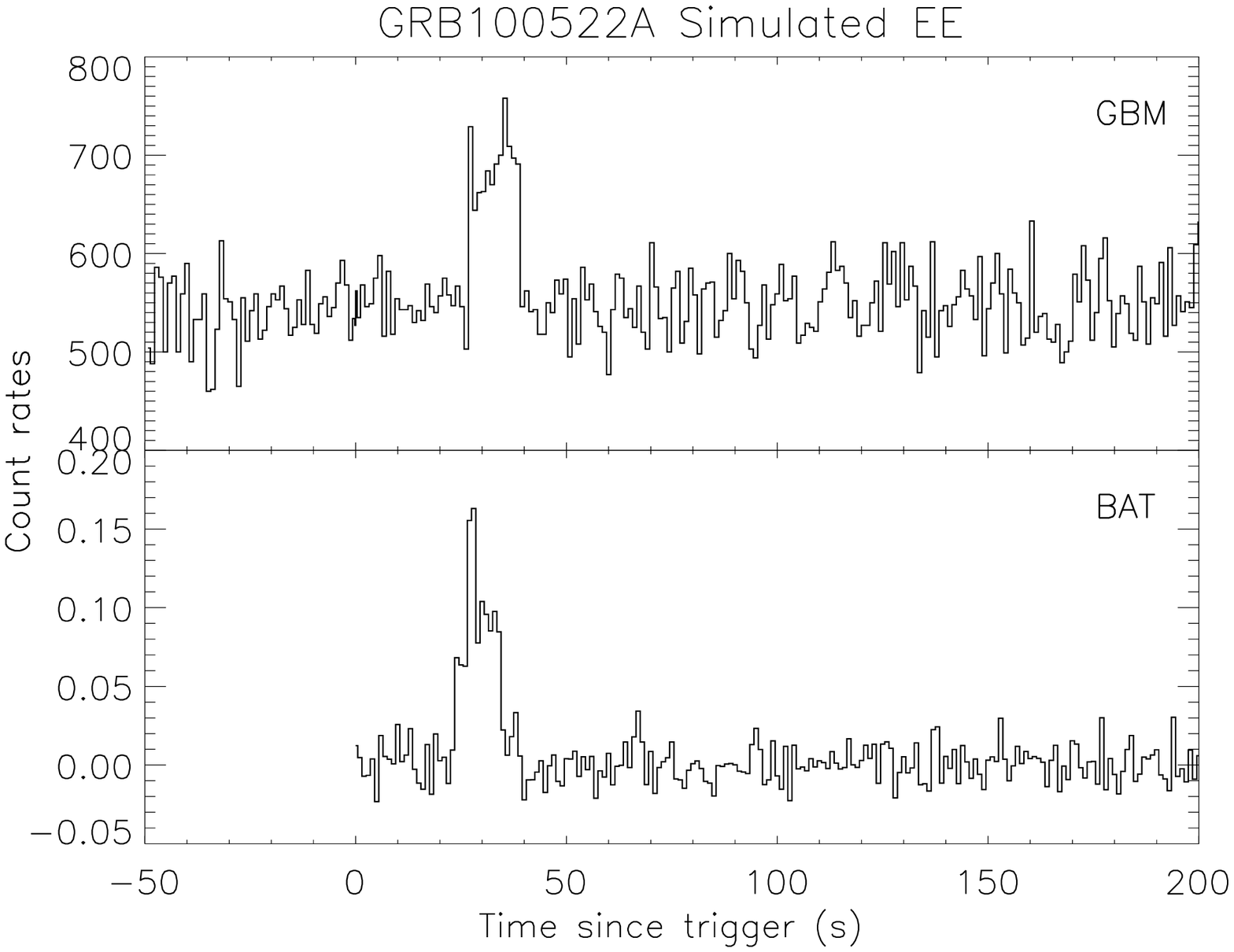}
\caption{[{\it Top}] Background lightcurve used in the simulations
for GRB 100522A, a common event observed with both GBM and BAT.
Here, we show the orbital background for GBM and the burst-free
lightcurve of the time interval after the actual detected EE, as
described in \S\ref{sec:sim}. [{\it Bottom}] Simulated
lightcurve for the same event, including noise-added EE. The EE
components shown here were simulated based on their Bayesian block
time profile and amplitude, in the energy range of actual EE
identification. }
  \label{100522a_sim}
\end{figure}

We show an example of simulated EE LCs and their background in
Figure~\ref{100522a_sim} for GRB\,100522A (a common event observed
with both GBM and BAT).  As seen in the figure, the simulated EE was
placed at the actual start time of the EE determined by the Bayesian
method (= $B_{\rm EE}$) for each instrument.

\subsubsection{ EE detection sensitivity}
 For each event and instrument, we varied
the amplitude of the simulated EE component nominally with
2$\times$, 1$\times$, 0.9$-$0.1$\times$ (in 0.1 steps), and
0$\times$ the actual Bayesian block amplitudes.  This was done for 3
energy bands: 15$-$25, 25$-$50, and 15$-$50 keV, one of which was
used for the original Bayesian block LC for each event. We simulated
10,000 LCs for each amplitude and energy band, and subjected them to
our EE search algorithm. For each set of LCs, our aim was to
determine the amplitude level at which at least 90\% of the
simulated LCs could be detected. To better constrain the 90
percentile amplitude level, we repeated the same procedure in
narrower amplitude interval, with finer amplitude steps as
necessary. We find that the 90 percentile amplitudes of both BAT and
GBM samples are anti-correlated with their corresponding flux, as
expected.

\begin{figure}
\centering
\includegraphics[trim=0cm 1cm 0cm 1cm, clip=true, scale=0.4]{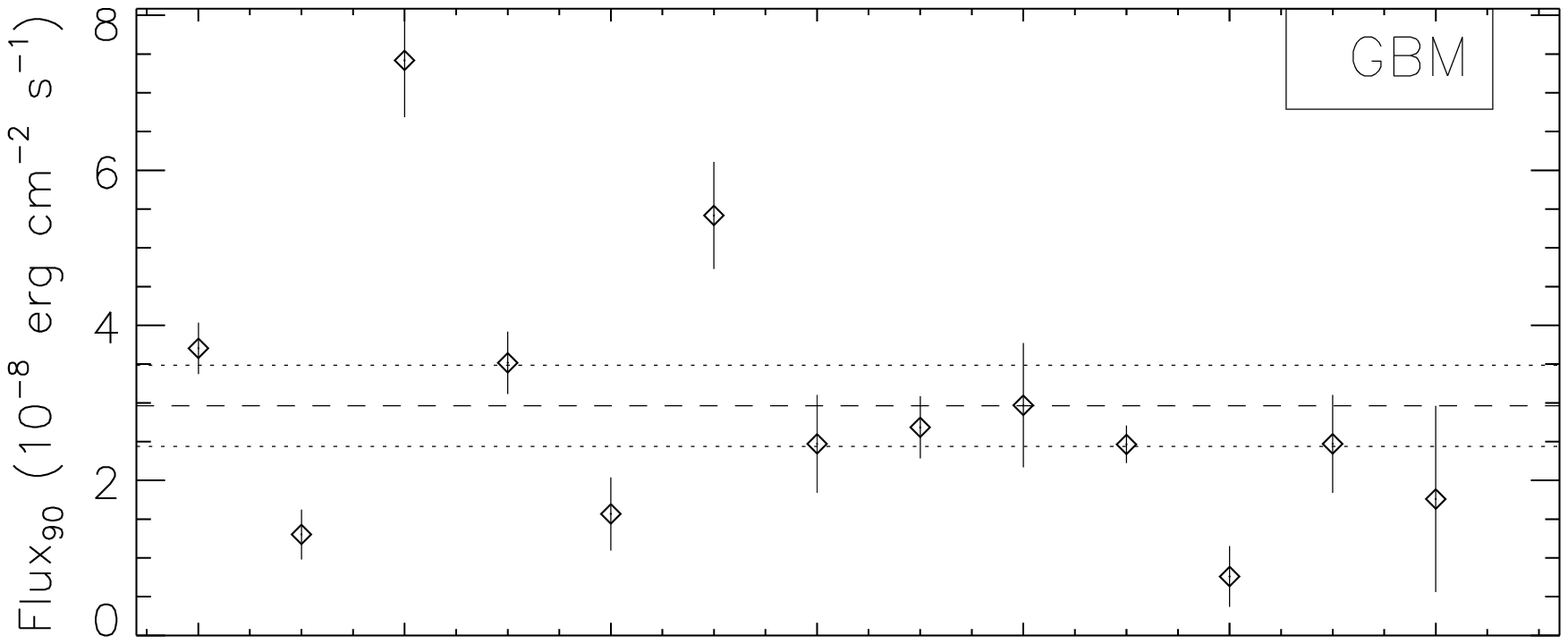}
\includegraphics[trim=0cm 1cm 0cm 1cm, clip=true, scale=0.4]{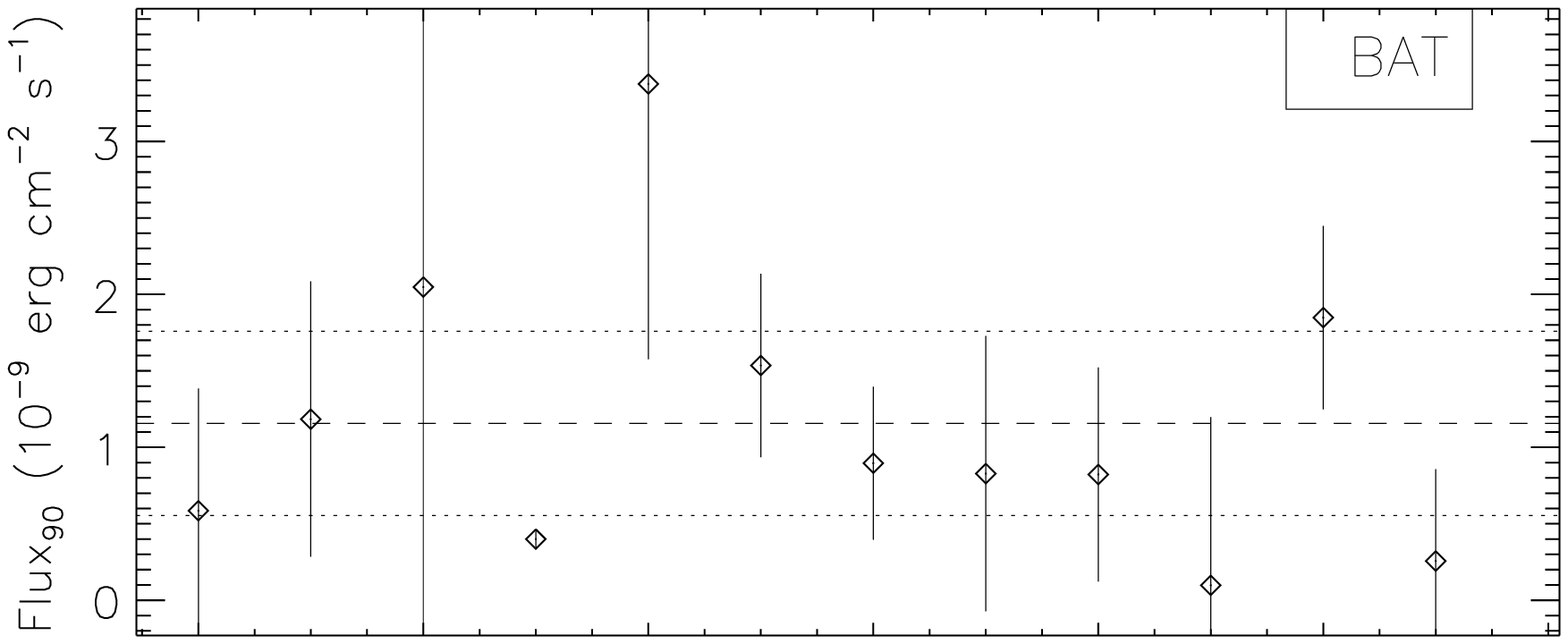}
\caption{The energy flux level for the 14 GBM events [{\it top}] and
16 BAT events [{\it bottom}] above which more than 90\% of the
simulated EE were detected.  The x-axis is the event numbers. The
average flux level of $2.9(3) \times 10^{-8}$\,ergs\,cm$^{-2}$
s$^{-1}$(GBM) and $1.2(6) \times 10^{-9}$\,ergs\,cm$^{-2}$ s$^{-1}$
(BAT) is indicated by the dashed line, with 1s standard deviation
(dotted lines).
 }
  \label{flux_limit}
\end{figure}

Finally, in order to determine the detection sensitivities while
minimizing the effect of various LC morphology on the detection
rates, we also simulated all EE with single step functions of the
actual EE durations and 10-$\sigma$ amplitudes.  Again, we simulated
10,000 LCs for the same set of amplitude values and energy bands as
above, and subjected them to our EE search algorithm.  Based on our
simulation results, we determined the 90 percentile EE detection
flux level (i.e., 90 percentile amplitude times the average EE flux)
for each event. We then determined the average 90 percentile flux
for each instrument. We find that the average energy flux level
above which more than 90\% of the simulated EE are detected is
$2.9(3) \times 10^{-8}$\,ergs\,cm$^{-2}$ s$^{-1}$ (for GBM) and
$1.2(6) \times 10^{-9}$\,ergs\,cm$^{-2}$ s$^{-1}$ (for BAT) in
15$-$50 keV (see Figure~\ref{flux_limit}).

\subsubsection{ False detection rates of our search
method}  As stated at the beginning of this section, the search
actually yielded more potential EE candidates than what we present
in this paper.  We went through the 5-panel lightcurves (e.g.,
Figure 2) one by one for these potential candidates, and excluded
the ones with obvious background issues, obvious long GRBs, or the
ones with EE seen only in single energy band. Then, our final list
of candidates included 16 BAT GRBs and 14 GBM GRBs. From these
numbers, we could say that the inferred false detection rates of our
search, with the specific set of criteria for the specific types of
EE, are 57\% and 79\%. Although these numbers may seem high, we
believe that the fact that we went through all the potential
candidate lightcurves and filtered out the actual falsely-detected
events, makes this not an issue for a concern.

Lastly, as an alternative way to probe the false detection rate due
to the data fluctuations and/or the choice of our background models,
we also searched for 12-s 1.5$\sigma$ deficit in the SNR LCs for all
events that were subjected to the original EE search. The search
identified such deficits in 6 BAT and 4 GBM events. Among the 6 BAT
deficits 4 of them are only seen in 50-100 keV band and well
isolated from the bursts by $>$150 s.  
The other 2 deficits were detected only in 25-50 keV band.
One of them starts at 230 s after the burst; however, the other one
(for GRB 121017A) was detected at $\sim$20 s after the burst, which
may have been identified as an EE candidate if this had been an
excess, and the indication of the deficit is also seen in 15--25
keV. Thus, the false detection rate based on the deficit search is
1/128 ($<$1\%) for the BAT sample. The deficit found for the 4 GBM
events were all clearly due to mismatched orbital background data.
Therefore, for our GBM sample, the false detection rate based on the
deficit search is even lower than that of the BAT sample.

\section{Properties of the EE Events}\label{sec:analysis}
\subsection{Spectral Lags}
One of the main characteristics that makes some GRBs with EE ``short''
GRBs is that the initial spikes usually show no (or close to nil)
spectral lags. Thus, we calculated the spectral lags for the initial
spikes of all of these events, as well as for the EE components where
possible. We used lightcurves with various time resolutions of 8 to
512 ms (for both BAT and GBM) in the energy bands of 50--100 keV and
15--25 keV for BAT, and 100--300 keV and 25--50\,keV for GBM, unless
otherwise noted. The time interval used for each event was the spike
or EE duration presented in Table~\ref{events}. For each event, we
determined a cross-correlation function between the lightcurves of the
two energy bands, which was subsequently fitted with a cubic function.
The peak of the fitted cubic function was taken as the lag. The
uncertainties in the lags were estimated with the same simulation
method that we used previously for the BATSE sample, which is
described in \citet{bos13}.
\begin{table}\centering
\setlength{\tabcolsep}{0.035in}
 \caption{Spectral lags calculated in 50--100\,keV/15--25\,keV for BAT
 and 100--300\,keV/25--50\,keV for GBM events unless otherwise noted}
    \begin{tabular}{lcccc}\hline
            &        \multicolumn{2}{c}{Spike}  &        \multicolumn{2}{c}{EE}  \\
 GRB name&  Lags   & Err          &     Lags   & Err          \\
                 &  (ms)    & (ms)       &     (ms)  &  (ms)         \\
    \hline
{\bf BAT}&               &              &                   &                   \\
050724  &    6.29    &   5.28    &      $-$     &  $-$       \\
051016B &    4.58    &   3.32    &      $-$     &  $-$      \\
060614  &    5.83    &   5.61    &  5.46    &   5.39         \\
061006  &    17.92   &   16.32   &  96.93   &   96.70        \\
061210  &    6.05    &   4.80    &      $-$      &  $-$     \\
070506  &    39.23   &   29.30   &  494.43  &  296.48$^*$    \\
070714B &    8.03    &   7.78    &      $-$     &  $-$    \\
080503  &   --7.56   &   4.82    &  40.51  &   8.95$^*$  \\
090531B &    5.21    &   4.07    &      $-$     &  $-$       \\
090927  &    4.77   &   2.38$^{**}$  &      $-$    &  $-$ \\
100212A &    2.53    &   2.42    &  255.14  &   204.83      \\
100522A &    26.36   &   25.77   &  78.48   &   64.86       \\
110207A &    3.15    &   2.78    &  38.33   &   37.34        \\
110402A &    32.93   &   28.64   &  40.39   &   37.43       \\
111121A &    1.00    &   0.93    &      $-$      &  $-$      \\
121014A & --17.30  &   13.59   &  596.32  &   560.79       \\
{\bf GBM}&                   &             &              &               \\
080807 & --3.38   &   3.54    &  46.18   &   44.18        \\
090131 &  101.84  &   95.49   &  89.10   &   79.18         \\
090820 & --0.01   &   0.49$^{**}$    &  196.86  &   131.38$^{**}$     \\
090831 &  28.97   &   19.88   &  23.72   &   28.57        \\
091120 &  57.50   &   51.34   &  41.98   &   42.35        \\
100517 &  76.22   &   62.79   &  -82.24  &   76.50        \\
100522 &  35.14   &   32.29   &      $-$   &  $-$        \\
110207 &  22.29   &   18.27   &  62.39   &   66.86        \\
110402 &  5.80    &   5.50    &  89.05   &   48.67         \\
110824 &  29.54   &   24.06   &  59.04   &   55.70         \\
111228 &  27.40   &   24.13   &      $-$    &  $-$        \\
120402 &  2.52    &   0.64    &      $-$      &  $-$      \\
120605 &  43.27   &   38.69   &  203.24  &   127.11$^{**}$     \\
121029 &  12.00   &   8.52    &  70.89   &   63.26         \\
\hline
  \end{tabular}
    \label{tab:lag}
\footnotesize{ \begin{flushleft}
$^*$ lag calculated in 25--50\,keV/15--25\,keV \\
$^{**}$ lag calculated in 50--100\,keV/25--50\,keV \\
\end{flushleft}}
\end{table}

\begin{figure}
\centering
\includegraphics[trim=0cm 0.5cm 0cm 0.9cm, clip=true, scale=0.4]{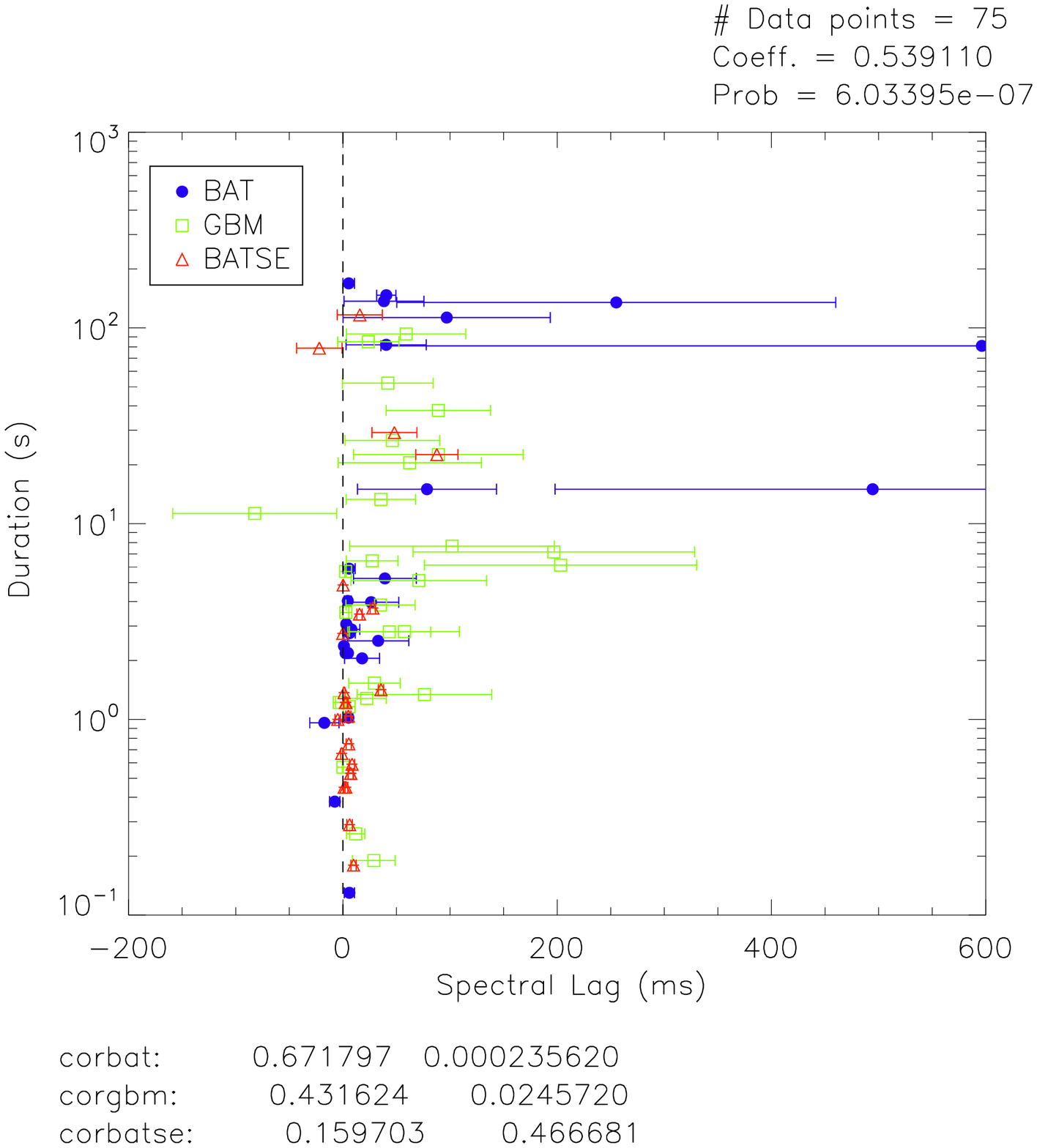}
\caption{Spectral lag vs.~duration ($T_{\rm spike}$ and $T_{\rm EE}$)
  of the 30 candidate events. BAT and GBM events are shown with
  different symbols and colors. For comparison, the values for BATSE
  GRBs with EE from our previous work are also
  included.}\label{lag}
\end{figure}

The spectral lag values that we found are presented in
Table~\ref{tab:lag}. For 20 events (9 BAT and 11 GBM), in which the
EE was bright enough to allow lag calculations, the lag values for
the EE are also shown in the table.  Although some of the errors are
large, the average lags are 19\,ms for the spikes and 122\,ms for
the extended emission. In Figure~\ref{lag}, we show spectral lags as
a function of their duration ($T_{\rm spike}$ or $T_{\rm EE}$).  For
a comparison, we also over-plot the values found from our previous
analysis of BATSE EE GRBs. There is an indication of a positive
correlation between the spectral lags and the duration ($T_{\rm
spike}$ or $T_{\rm EE}$) within the BAT sample as well as when all
samples are combined (BAT + GBM + BATSE; see
\S\ref{sec:correlation}, albeit large uncertainties associated with
some values. We also looked at the correlations with $z$-corrected
(source frame) durations for the 8 events with known redshift values
but found null results.

\subsection{Spectral Analysis}\label{sec:spec}
In order to examine spectral properties of these short GRBs with EE
as well as to obtain their energetics, we performed spectral
analysis of these events using RMFIT\footnote[2]{R. S. Mallozzi, R.
D. Preece, \& M. S. Briggs,
  ''RMFIT, A Lightcurve and Spectral Analysis Tool,'' \copyright~
  Robert D. Preece, University of Alabama in Huntsville.} version
  4.0rc1.
For each event, we analyzed the spectra of the spike and the EE
separately. The time intervals of the spike and EE spectra were matched
to those given in Table~\ref{events}.

For the BAT events, we extracted spectra with an energy range of
15$-$150\,keV (80 energy channels) and generated corresponding
detector response matrices using {\tt batdrmgen}, which utilizes the
online calibration database. For the GBM events, we used
 TTE data providing 128 energy channels in
$\sim$5--2000\,keV binned to 64-ms time resolution.  The actual
fitted energy range was $\sim$8--1000\,keV after excluding several
low-energy channels and the high-energy overflow channel
\citep{gru14}. For each event, we selected a set of
 NaI detectors with detector-to-source angles less than $60^o$
\citep{pac12}, and fit the data of these detectors simultaneously.
Since some of the EE components were rather dim, the accurate
background modeling was essential for the spectral analysis also. We
modeled the background spectrum of each of the GBM event with a
polynomial function, and subsequently compared it to match the orbital
background we used in the search procedure.  To account for a possible
systematic normalization offset between the multiple detectors, we
also included multiplicative ``effective area correction" factor (EAC)
in the initial fits using spike spectra of the bursts. We did find,
however, in each case, the normalization factors were very close to 1
in most cases and were fixed in the final fits.

We fitted each spectrum (spike and EE separately) with three models:
a power law (PWRL), cut-off power law (or ``Comptonized"; COMP), and
the Band function \citep[BAND;][]{band93}, by minimizing $\chi^2$.
The best-fit model for each spectrum was subsequently determined
based on the improvements in their $\chi^2$ values for additional
degrees of freedom. In Table~\ref{specanlysis}, we present the
best-fit model parameters along with the energy fluence, peak flux,
and hardness ratios. The hardness ratio here was calculated in
photon space using the best-fit spectral parameters.
\begin{table*}
\center \scriptsize \setlength{\tabcolsep}{0.025in} \caption{Summary
of spectral fit results of GRBs with EE. }
\begin{tabular}{lclcccccc}
\hline
    GRB  &  Component & Model  &  $\lambda$ & $E\rm{_{peak}}$&$\chi^2$/dof &Energy$\rm{^{b}}$ &${F\rm{_{peak}}}^{\rm c}$  & Hardness \\
    Name &            &        &        & (keV)      &         &Fluence       &   &  Ratio $^{\rm d}$        \\
\hline
{\bf BAT}&           &         &    &        &         &       &     &      \\
 050724  &   Spike   & PWRL    &  --1.67 $\pm$ 0.12  &       & 62.5  / 56  &  4.3   $\pm$ 0.2    &   6.8  $\pm$ 0.7  &   1.25 $\pm$ 0.14    \\
         &   EE      & PWRL    &  --2.04 $\pm$ 0.18  &       & 71.7  / 56  &  7.3   $\pm$ 0.9    &   0.3  $\pm$ 0.3  &   0.98 $\pm$ 0.17    \\
 051016B &   Spike   & PWRL    &  --2.42 $\pm$ 0.17  &       & 72.8  / 56  &  1.8   $\pm$ 0.2    &   1.1  $\pm$ 0.3  &   0.75 $\pm$ 0.14    \\
         &   EE      & PWRL    &  --2.87 $\pm$ 0.57  &       & 47.1  / 56  &  1.2   $\pm$ 0.3    &   0.3  $\pm$ 0.2  &   0.55 $\pm$ 0.38    \\
 060614  &   Spike   & PWRL    &  --1.67 $\pm$ 0.04  &       & 46.6  / 56  &  40.4  $\pm$ 1.1    &   15.5 $\pm$ 1.7  &   1.26 $\pm$ 0.05    \\
         &   EE      & PWRL    &  --2.10 $\pm$ 0.02  &       & 55.0  / 56  &  115.3 $\pm$ 1.2    &   4.0  $\pm$ 1.0  &   0.93 $\pm$ 0.02    \\
 061006  &   Spike   & PWRL    &  --1.07 $\pm$ 0.06  &       & 38.0  / 56  &  7.4   $\pm$ 0.3    &   12.3 $\pm$ 0.8  &   1.89 $\pm$ 0.09    \\
         &   EE      & PWRL    &  --2.23 $\pm$ 0.16  &       & 54.5  / 56  &  8.6   $\pm$ 0.8    &   0.5  $\pm$ 0.3  &   0.85 $\pm$ 0.14    \\
 061210  &   Spike   & PWRL    &  --0.75 $\pm$ 0.14  &       & 34.9  / 56  &  4.2   $\pm$ 0.4    &   24.9 $\pm$ 1.7  &   2.39 $\pm$ 0.31    \\
         &EE$^{*}$   & PWRL    &  --1.73 $\pm$ 0.20  &       & 85.0  / 56  &  9.2   $\pm$ 1.2    &   ---  &   1.21 $\pm$ 0.23    \\
 070506  &   Spike   & PWRL    &  --1.79 $\pm$ 0.12  &       & 52.4  / 56  &  2.6   $\pm$ 0.2    &   0.5  $\pm$ 0.2  &   1.14 $\pm$ 0.15    \\
         &   EE      & PWRL    &  --2.88 $\pm$ 0.71  &       & 40.4  / 56  &  0.6   $\pm$ 0.2    &   0.4  $\pm$ 0.3  &   0.53 $\pm$ 0.45    \\
 070714B &   Spike   & PWRL    &  --1.10 $\pm$ 0.07  &       & 51.7  / 56  &  6.8   $\pm$ 0.3    &   5.7  $\pm$ 0.6  &   1.87 $\pm$ 0.11    \\
         &   EE      & PWRL    &  --2.33 $\pm$ 0.34  &       & 40.5  / 56  &  1.8   $\pm$ 0.4    &   0.3  $\pm$ 0.2  &   0.80 $\pm$ 0.30    \\
 080503  &   Spike   & PWRL    &  --1.85 $\pm$ 0.39  &       & 61.9  / 56  &  0.6   $\pm$ 0.2    &   1.8  $\pm$ 0.4  &   1.07 $\pm$ 0.40    \\
         &   EE      & PWRL    &  --1.93 $\pm$ 0.08  &       & 44.2  / 56  &  19.4  $\pm$ 0.9    &   0.1  $\pm$ 0.1  &   1.06 $\pm$ 0.08    \\
 090531B &   Spike   & PWRL    &  --1.25 $\pm$ 0.12  &       & 58.4  / 56  &  2.4   $\pm$ 0.2    &   2.1  $\pm$ 0.4  &   1.68 $\pm$ 0.17    \\
         &   EE      & PWRL    &  --1.79 $\pm$ 0.17  &       & 54.1  / 56  &  5.0   $\pm$ 0.1    &   0.5  $\pm$ 0.3  &   1.16 $\pm$ 0.19    \\
 090927  &   Spike   & PWRL    &  --1.88 $\pm$ 0.16  &       & 54.9  / 56  &  2.2   $\pm$ 0.3    &   2.1  $\pm$ 0.5  &   1.09 $\pm$ 0.18    \\
         &   EE      & PWRL    &  --1.98 $\pm$ 0.45  &       & 67.5  / 56  &  1.9   $\pm$ 0.6    &   0.8  $\pm$ 0.4  &   1.00 $\pm$ 0.46    \\
 100212A &   Spike   & PWRL    &  --1.42 $\pm$ 0.10  &       & 53.3  / 56  &  3.1   $\pm$ 0.2    &   2.0  $\pm$ 0.4  &   1.50 $\pm$ 0.13    \\
         &   EE      & PWRL    &  --2.47 $\pm$ 0.21  &       & 64.3  / 56  &  6.1   $\pm$ 0.8    &   0.6  $\pm$ 0.3  &   0.73 $\pm$ 0.24    \\
 100522A &   Spike   & PWRL    &  --1.67 $\pm$ 0.04  &       & 45.5  / 56  &  17.0  $\pm$ 0.5    &   6.9  $\pm$ 0.7  &   1.26 $\pm$ 0.05    \\
         &   EE      & PWRL    &  --2.56 $\pm$ 0.10  &       & 45.3  / 56  &  6.1   $\pm$ 0.3    &   0.6  $\pm$ 0.3  &   0.68 $\pm$ 0.07    \\
 110207A &   Spike   & PWRL    &  --1.29 $\pm$ 0.14  &       & 46.6  / 56  &  2.5   $\pm$ 0.3    &   6.6  $\pm$ 0.8  &   1.63 $\pm$ 0.21    \\
         &   EE      & PWRL    &  --1.46 $\pm$ 0.09  &       & 48.2  / 56  &  17.0  $\pm$ 1.1    &   1.3  $\pm$ 0.4  &   1.45 $\pm$ 0.12    \\
 110402A &   Spike   & PWRL    &  --1.04 $\pm$ 0.15  &       & 55.7  / 56  &  9.9   $\pm$ 0.9    &   10.5 $\pm$ 2.1  &   1.94 $\pm$ 0.27    \\
         &   EE      & PWRL    &  --1.93 $\pm$ 0.14  &       & 52.7  / 56  &  29.7  $\pm$ 2.7    &   2.7  $\pm$ 1.1  &   1.05 $\pm$ 0.14    \\
 111121A &   Spike   & PWRL    &  --1.22 $\pm$ 0.06  &       & 67.2  / 56  &  10.9  $\pm$ 0.4    &   16.0 $\pm$ 1.3  &   1.72 $\pm$ 0.09    \\
         &   EE      & PWRL    &  --2.02 $\pm$ 0.10  &       & 58.6  / 56  &  11.8  $\pm$ 0.7    &   0.6  $\pm$ 0.3  &   1.02 $\pm$ 0.10    \\
 121014A &   Spike   & PWRL    &  --2.27 $\pm$ 0.16  &       & 60.8  / 56  &  1.5   $\pm$ 0.2    &   2.1  $\pm$ 0.4  &   0.83 $\pm$ 0.15    \\
         &   EE      & PWRL    &  --2.00 $\pm$ 0.14  &       & 51.4  / 56  &  9.7   $\pm$ 0.8    &   1.2  $\pm$ 0.3  &   1.00 $\pm$ 0.13    \\

{\bf GBM} &       &         &        &    & &   &     &      \\
 080807&  Spike   & PWRL    &  --1.08 $\pm$ 0.02  &        & 487.7 / 480 &  11.1 $\pm$ 0.3  &   23.0 $\pm$ 2.3  &   3.72 $\pm$ 0.16    \\
          &   EE     & PWRL &  --1.33 $\pm$ 0.04  &        & 523.9 / 480 &  18.5 $\pm$ 0.9  &   9.5  $\pm$ 1.8  &   2.91 $\pm$ 0.18    \\
 090131&  Spike   & COMP    &  --1.16 $\pm$ 0.04  & 56.7 $\pm$ 1.2 & 743.9 / 476 &  30.5 $\pm$ 0.4  &   40.6 $\pm$ 3.0  &   0.77 $\pm$ 0.04    \\
          &   EE     & COMP &  --1.49 $\pm$ 0.03  & 154.5$\pm$ 14.0& 642.7 / 476 &  49.6 $\pm$ 0.7  &   48.8 $\pm$ 3.8  &   1.74 $\pm$ 0.05    \\
 090820&  Spike   & PWRL    &  --1.68 $\pm$ 0.04  &        & 377.4 / 356 &   4.3 $\pm$ 0.2  &   9.0  $\pm$ 1.7  &   2.11 $\pm$ 0.16    \\
         &   EE      & PWRL &  --2.22 $\pm$ 0.06  &        & 409.5 / 356 &   7.2 $\pm$ 0.5  &   9.7  $\pm$ 1.8  &   1.30 $\pm$ 0.20    \\
 090831&  Spike   & PWRL    &  --1.34 $\pm$ 0.03  &        & 258.1 / 235 &   7.1 $\pm$ 0.3  &   51.6 $\pm$ 3.3  &   2.91 $\pm$ 0.16    \\
         &   EE      & COMP &  --1.65 $\pm$ 0.06  & 532  $\pm$ 422 & 218.5 / 234 & 112.2 $\pm$ 3.4  &   24.0 $\pm$ 3.6  &   2.03 $\pm$ 0.12    \\
 091120&  Spike   & COMP    &  --0.76 $\pm$ 0.05  & 277.7$\pm$ 18.7& 515.3 / 470 &  35.1 $\pm$ 0.5  &   28.5 $\pm$ 3.7  &   2.93 $\pm$ 0.08    \\
         &   EE      & COMP &  --1.10 $\pm$ 0.03  & 114.4$\pm$ 3.9 & 631.8 / 470 & 192.3 $\pm$ 2.5  &   37.9 $\pm$ 4.2  &   1.52 $\pm$ 0.04    \\
 100517&  Spike   & COMP    &  --0.69 $\pm$ 0.14  & 92.7 $\pm$ 6.8 & 328.2 / 354 &   9.2 $\pm$ 0.4  &   22.7 $\pm$ 3.2  &   0.05 $\pm$ 0.02    \\
         &   EE      & COMP &  --1.39 $\pm$ 0.19  & 26.8 $\pm$ 3.2 & 367.2 / 354 &  14.0 $\pm$ 0.8  &   6.5  $\pm$ 1.5  &   0.36 $\pm$ 0.12    \\
 100522&  Spike   & COMP    &  --0.91 $\pm$ 0.07  & 143.8$\pm$10.2 & 445.7 / 470 &  24.7 $\pm$ 0.6  &   18.9 $\pm$ 2.7  &   1.83 $\pm$ 0.09    \\
         &   EE      & PWRL &  --2.21 $\pm$ 0.07  &        & 466.3 / 471 &   9.9 $\pm$ 0.8  &   4.7  $\pm$ 1.5  &   1.32 $\pm$ 0.23    \\
 110207&  Spike   & PWRL    &  --1.25 $\pm$ 0.05  &        & 363.1 / 351 &   5.1 $\pm$ 0.3  &   14.2 $\pm$ 2.0  &   3.54 $\pm$ 0.38    \\
         &   EE      & PWRL &  --1.23 $\pm$ 0.06  &        & 353.1 / 351 &  22.3 $\pm$ 1.5  &   10.3 $\pm$ 1.9  &   3.24 $\pm$ 0.33    \\
 110402&  Spike   & PWRL    &  --1.18 $\pm$ 0.09  &        & 232.5 / 235 &   4.2 $\pm$ 0.5  &   34.3 $\pm$ 3.5  &   3.38 $\pm$ 0.59    \\
         &   EE      & PWRL &  --1.43 $\pm$ 0.05  &        & 265.6 / 235 &  58.5 $\pm$ 3.1  &   9.3  $\pm$ 2.4  &   2.68 $\pm$ 0.22    \\
 110824&  Spike   & PWRL    &  --1.02 $\pm$ 0.02  &        & 493.4 / 475 &  23.3 $\pm$ 0.4  &   41.2 $\pm$ 2.8  &   3.92 $\pm$ 0.12    \\
         &   EE      & PWRL &  --1.59 $\pm$ 0.03  &        & 541.2 / 475 &  69.4 $\pm$ 2.2  &   25.7 $\pm$ 2.4  &   2.30 $\pm$ 0.12    \\
 111228&  Spike   & COMP    &  --0.99 $\pm$ 0.08  & 33.9 $\pm$ 0.9 & 476.8 / 475 &  25.7 $\pm$ 0.5  &   18.9 $\pm$ 2.4  &   0.30 $\pm$ 0.04    \\
         &EE$^{**}$ & ~~~-  &  ---        &        & ---     &    ---       &   ---     &   ----    \\
 120402&  Spike   & COMP    &  --1.16 $\pm$ 0.08  & 50.2 $\pm$ 2.2 & 236.8 / 235 &  22.8 $\pm$ 0.7  &   15.2 $\pm$ 2.3  &   0.66 $\pm$ 0.08    \\
         &   EE      & PWRL &  --2.12 $\pm$ 0.13  &        & 293.8 / 235 &   4.7 $\pm$ 0.8  &   5.2  $\pm$ 1.7  &   1.35 $\pm$ 0.45    \\
 120605&  Spike   & COMP    &  --1.13 $\pm$ 0.07  & 375.7$\pm$80.1 & 403.9 / 351 &  16.7 $\pm$ 0.5  &   28.4 $\pm$ 3.7  &   2.70 $\pm$ 0.14    \\
         &   EE      & PWRL &  --1.98 $\pm$ 0.06  &        & 338.7 / 352 &   6.6 $\pm$ 0.5  &   6.4  $\pm$ 1.6  &   1.62 $\pm$ 0.23    \\
 121029&  Spike   & COMP    &  --0.74 $\pm$ 0.15  & 338.3$\pm$68.7 & 545.7 / 561 &   4.4 $\pm$ 0.2  &   19.7 $\pm$ 2.8  &   3.25 $\pm$ 0.26    \\
         &   EE      & COMP &  --0.34 $\pm$ 0.06  & 178.6$\pm$ 6.1 & 577.7 / 561 &  52.6 $\pm$ 0.7  &   40.4 $\pm$ 4.3  &   2.48 $\pm$ 0.06    \\

\hline
\end{tabular}
 \footnotesize{ \begin{flushleft}  All uncertainties are 1$\sigma$.\\
$\rm{^{a}}$ in units of $\rm{10^{-5}~photon~cm^{-2}~s^{-2}~keV^{-1}}$ \\
$\rm{^{b}}$ in units of $\rm {10^{-7}~erg~cm^{-2}}$ and calculated in the 15$-$350~keV range \\
$\rm{^{c}}$ in units of $\rm {10^{-7}~erg~cm^{-2}~s^{-1}}$ and calculated in the 15$-$350~keV range with 64-ms resolution \\
$\rm{^{d}}$ calculated in the 25$-$50~keV and 50$-$100~keV ranges for BAT and the 50$-$100~keV and 100$-$300~keV ranges for GBM\\
${^{*}}$ no enough statistic for F$\rm{_{peak}}$ calculation.\\
${^{**}}$ could not perform the spectral fit due to its weak nature
\end{flushleft}}
\label{specanlysis}
\end{table*}

Both the spikes and the EE of all of the BAT events were
sufficiently described with PWRL, as expected for the narrow-band
spectra of BAT.  On the other hand, some GBM spectra are described
better with COMP model.

Interestingly, we noticed that the BAT lightcurve of GRB\,100212A
displayed a secondary spike (at $T$+80\,s, see
Figure~\ref{common_lc}) dimmer than the initial spike at $T_0$,
followed by secondary EE. This second episode (both spike and the
EE) is actually seen as a X-ray flare in {\it Swift} XRT data of the
same event\footnotemark[3] \footnotetext[3]{The XRT lightcurve is
available at http://www.swift.ac.uk/xrt\_curves/00412081/}.  We
attempted to extract the spectra of the secondary spike and the EE
that follows using BAT data (as well as BAT+GBM), but the component
was too weak for a spectral analysis.  Nevertheless, it is obvious
from these lightcurves that the second episode is much softer than
the first spike and the first EE (at $\sim T$+10\,s).  For the
spectral analysis of EE of this event presented in
Table~\ref{specanlysis}, the duration used as the EE includes the
first EE and the entire second episode.

\subsection{Common Events}\label{sec:common}
\begin{figure*}
\centering
\includegraphics[height=10cm]{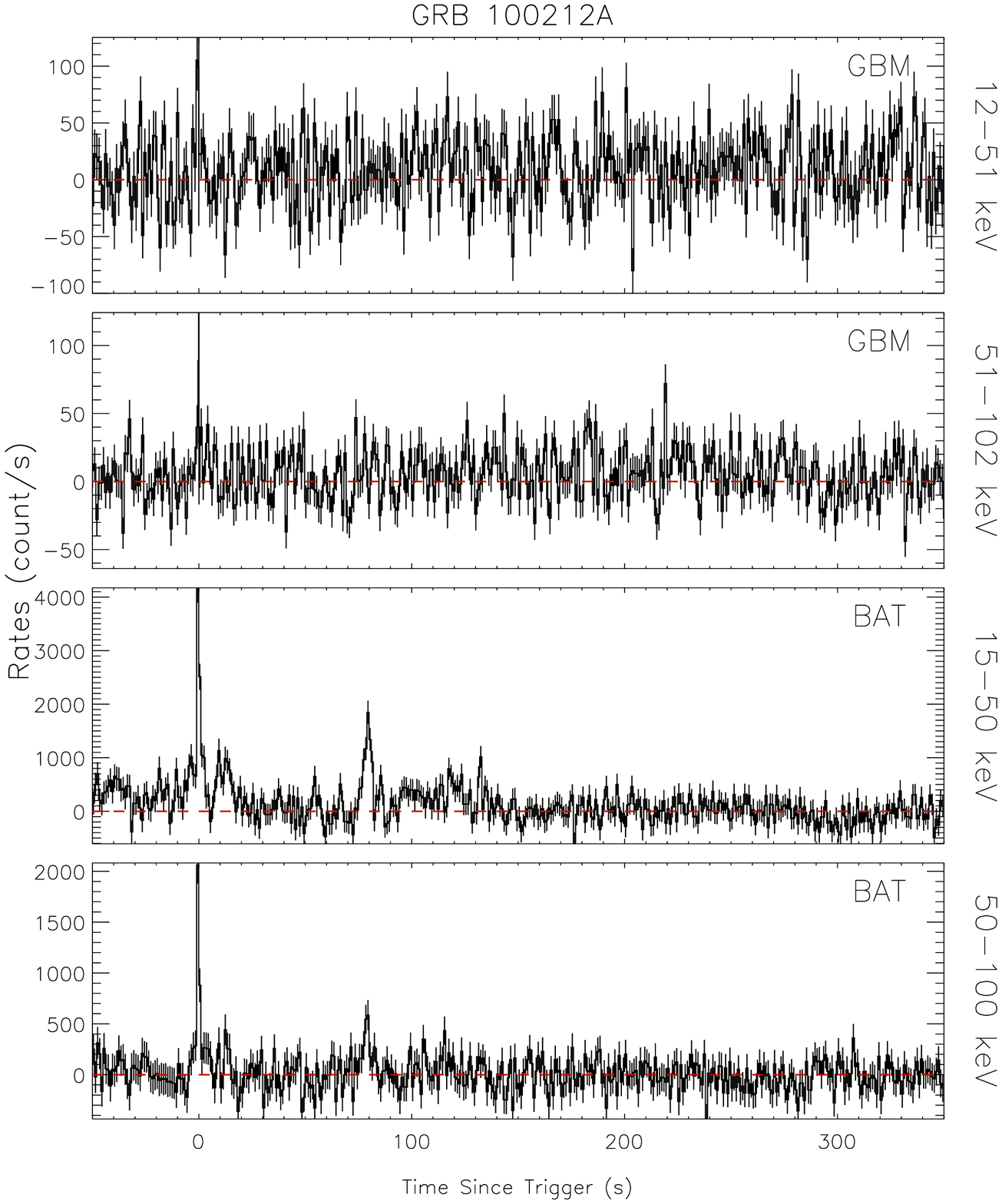}
\includegraphics[height=10cm]{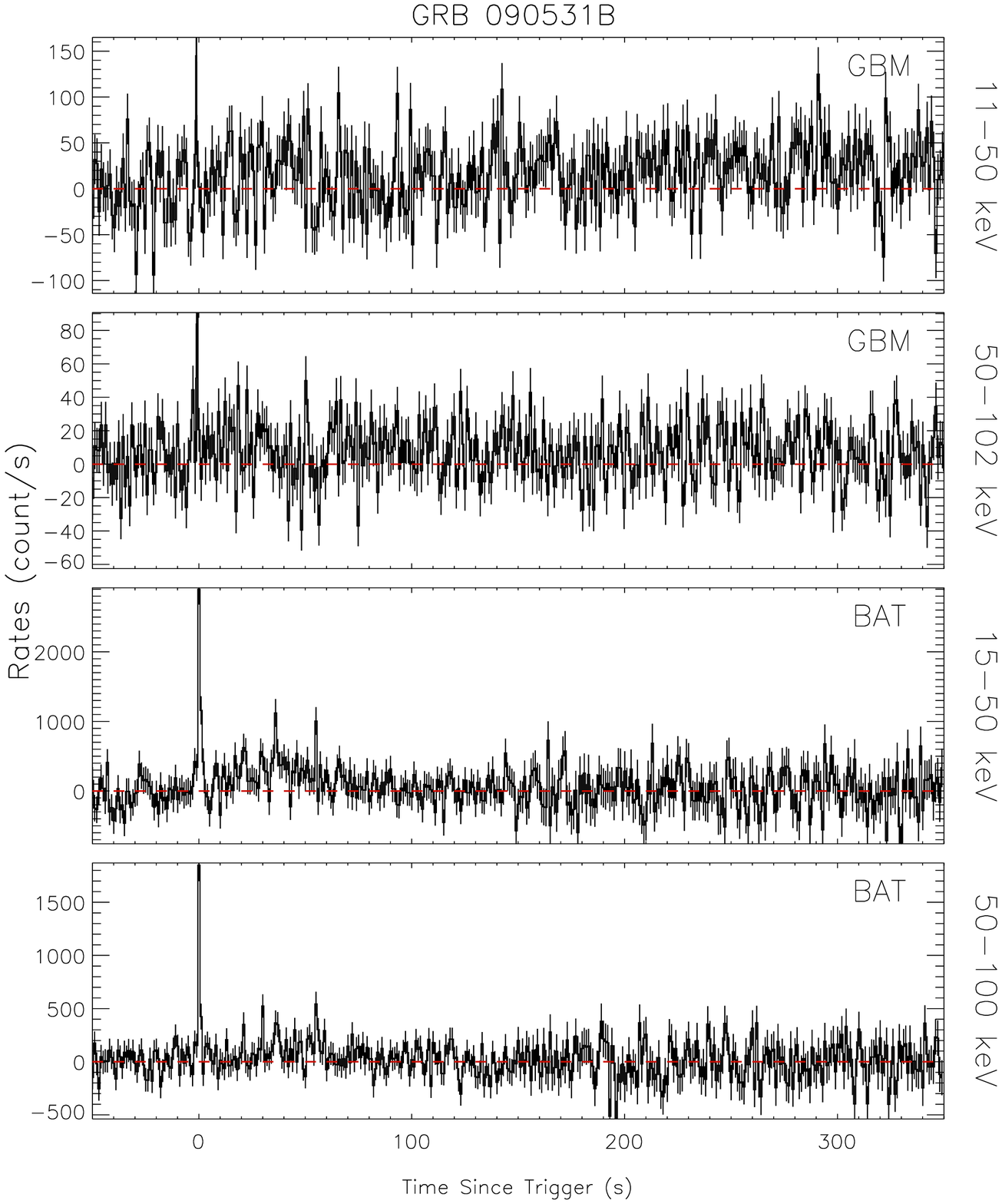}
\caption{The background-subtracted lightcurves of GRBs\,100212A and
  090531B observed both with BAT (bottom two panels) and with GBM
  (top two panels).  The EE was identified
  only in the BAT data.  } \label{common_lc}
\end{figure*}

There were a total of 5 GRBs with EE that were observed with both
BAT and GBM.  For 3 of them EE was identified in both BAT and GBM
data in our search (see Table~\ref{common} for a summary). The BAT
and GBM lightcurves of two example common events are shown in
Figure~\ref{common_lc}.
\begin{table*}
\caption{List of common events observed with both BAT and GBM.}
    \begin{tabular}{lcc}
    \hline
  &   \multicolumn{2}{c}{EE detected in the search}    \\
GRB name    &   in BAT   &   in GBM      \\ \hline
090531B &   yes &   no but visible in LC       \\
100212A &   yes &   no         \\
100522A &   yes &   yes        \\
110207A &   yes &   yes   \\
110402A &   yes &   yes        \\ \hline
    \end{tabular}
    \label{common}
\end{table*}
%


%

The EE of these two events were identified only in BAT data.
Generally speaking, non-detection of EE by GBM may
naturally be expected for some weak, soft EE, since BAT has larger
effective area over the lower energy range of 10--100\,keV than GBM
NaI detectors \citep{sta09}.

\begin{figure*}
\centering
\includegraphics[clip=true,scale=0.4]{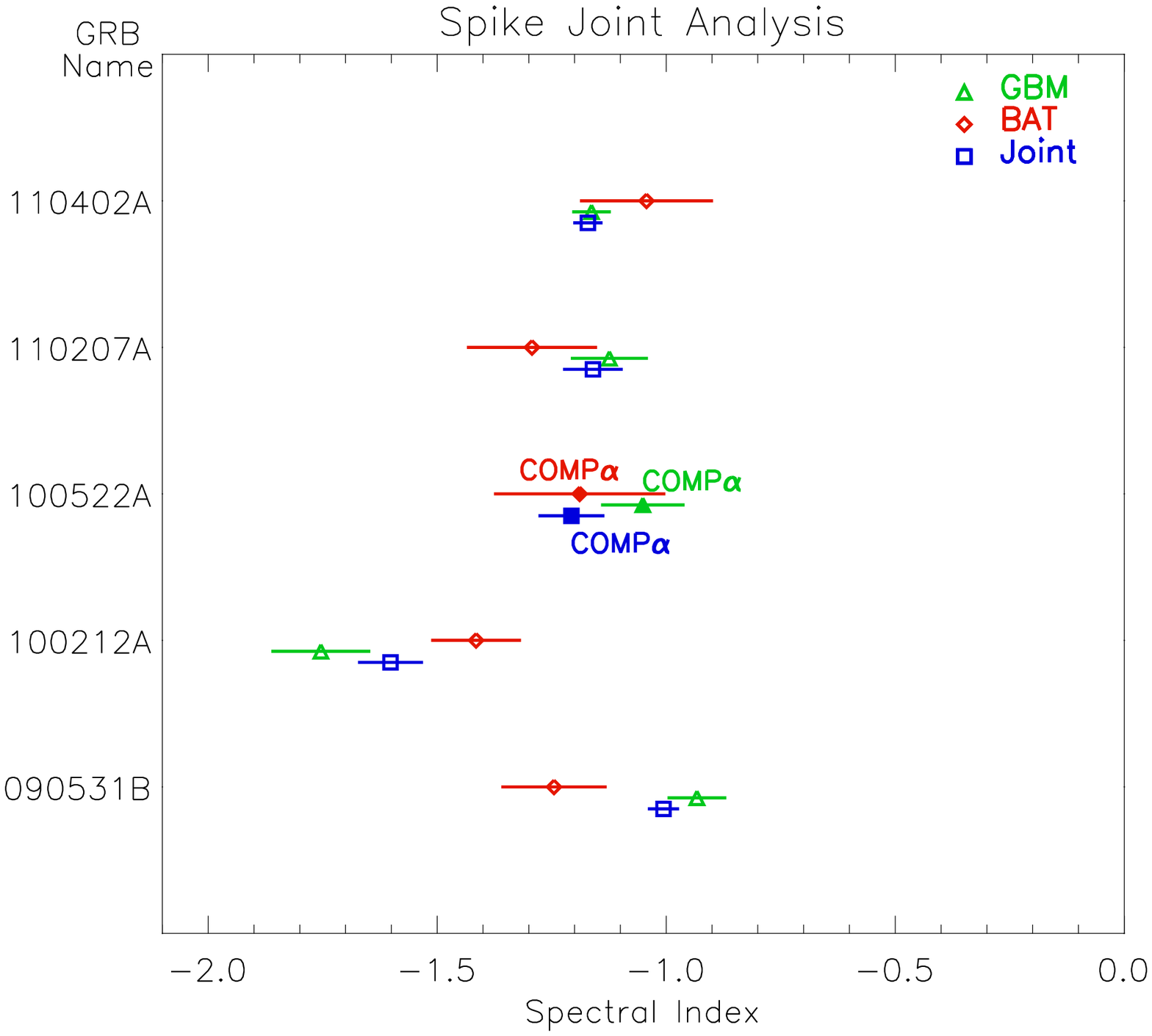}
\includegraphics[clip=true,scale=0.4]{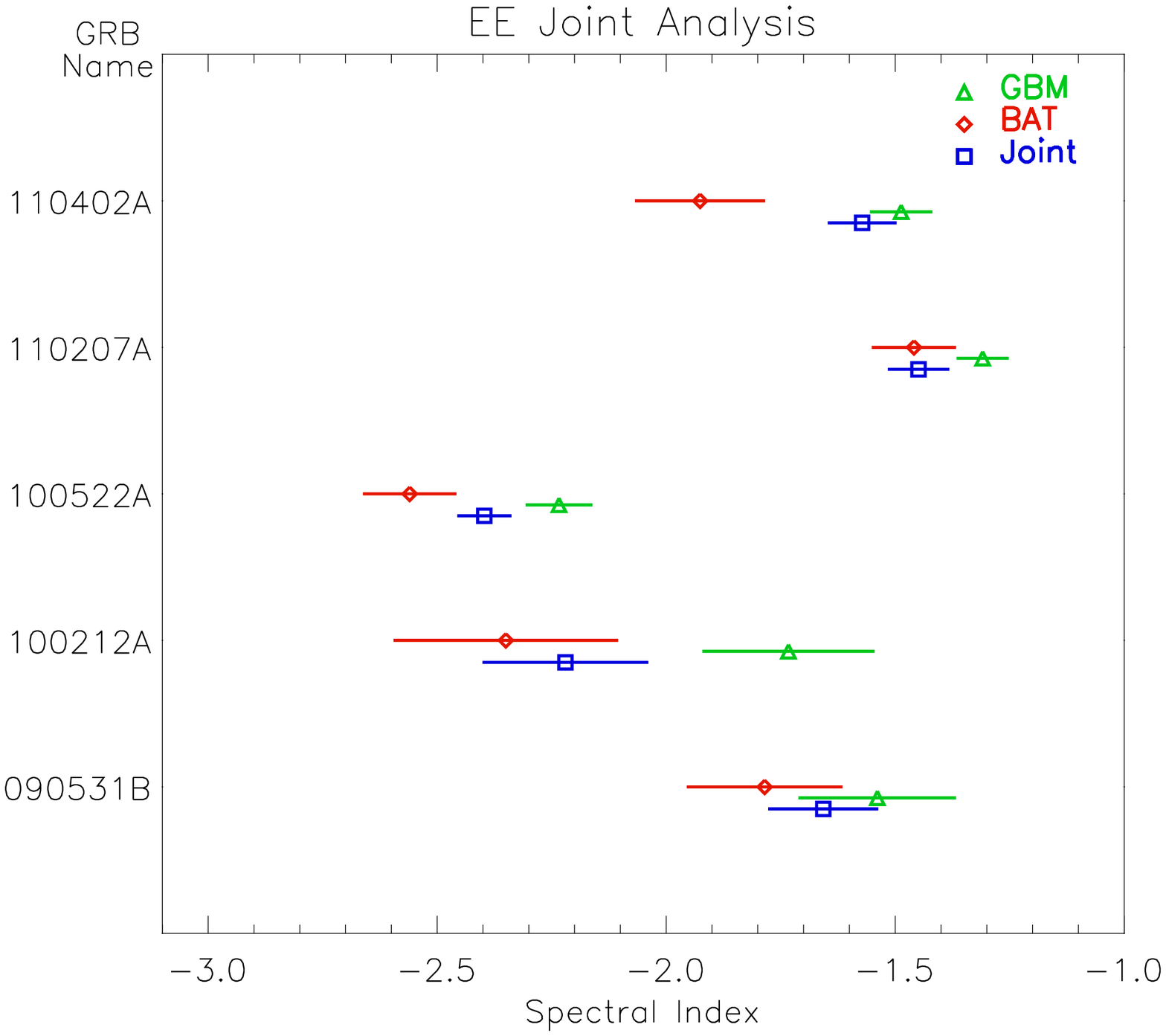}
\caption{The best-fit PWRL indices of the spikes (left) and EE
(right) of the 5 common events observed with both BAT and GBM. In
the cases where COMP provides better fits, the low-energy
indices ($\alpha$) are also shown.  The indices found in single
detector analysis (either BAT or GBM) and in joint analysis
(BAT+GBM) are shown for each event. } \label{joint_ind}
\end{figure*}
For these 5 common events, we performed joint spectral analysis
using both BAT and GBM data of the spikes and the EE.
 The time intervals of the joint analysis were
chosen considering the start and end times of each component of both
BAT and GBM data.  Then, the earlier start time and later end time
of the two were taken as the joint analysis time interval. The exact
time intervals are limited by the resolution of GBM data; In order
to match the time intervals of both spectra analyzed jointly, we
extracted BAT spectra of all these common events with the selected
time intervals of the GBM spectra. Moreover, we initially included
the EAC factors in the joint fits to account for possible systematic
discrepancy between the BAT and GBM NaI data; however, the factors
were always close to unity and therefore, not needed in the final
fits.

In Figure~\ref{joint_ind}, we compare the spectral indices found in
the joint analysis with those found in the single-detector analysis
(BAT-only and GBM-only). Most of the joint spectra of both spikes
and EE were still best described by PWRL, and there was one case
(100522A) in which COMP was better fit to the spectra. In that case,
the low-energy power-law index ($\alpha$) values are shown in the
Figure and indicated as such. Although the BAT energy range used for
the spectral analysis (15$-$150\,keV) lies well within the GBM
energy range (8$-$1000\,keV), the joint analysis constrained the
parameters better in almost all cases as seen in the Figure. The
parameters found in the joint analysis of the spikes and EE are more
consistent with those found in the GBM-only and BAT-only analysis
respectively, although the GBM and BAT parameters are usually
consistent within 1--2$\sigma$ (if the same models are used). The
joint analysis illustrates that using only the BAT spectrum is not
always sufficient to determine the real spectral shape due to the
its narrow coverage in relatively softer energy range; nonetheless,
adding the BAT data to broader GBM spectra can help better constrain
the spectral parameters.

\subsection{Correlations}\label{sec:correlation}
To study the general properties of the spikes and the EE of the 30
candidate events, we looked for correlations among their spectral
parameters, durations, lags, and flux/fluence.  The correlations
were searched separately for the BAT and GBM samples to reduce the
chance of finding a false correlation caused by possible systematic
differences between the samples.  Moreover, we studied the
correlations in several different sampling: within the spikes of the
30 events, within the EE components, as well as spike vs.~EE, and
the spikes and EE all together.

The exact parameters used for the correlation study are: burst's
duration ($T_{90}$), durations of each component ($T_{\rm spike}$
and $T_{\rm EE}$), spectral lag, best-fit spectral parameters,
energy fluence, peak energy flux, hardness ratio, and total burst energy
fluence.  For the 8 events with $z$ measurements, we also looked at
their isotropic-equivalent total energy ($E_{\rm iso}$), luminosity,
and source-frame durations (i.e., $T/(1+z)$).  We also compared the
parameters we found in this study to the BATSE sample from our
previous study \citep{bos13}.

\begin{figure}
\centering
\includegraphics[trim=0cm 0cm 0cm 0.9cm, clip=true, scale=0.4]
{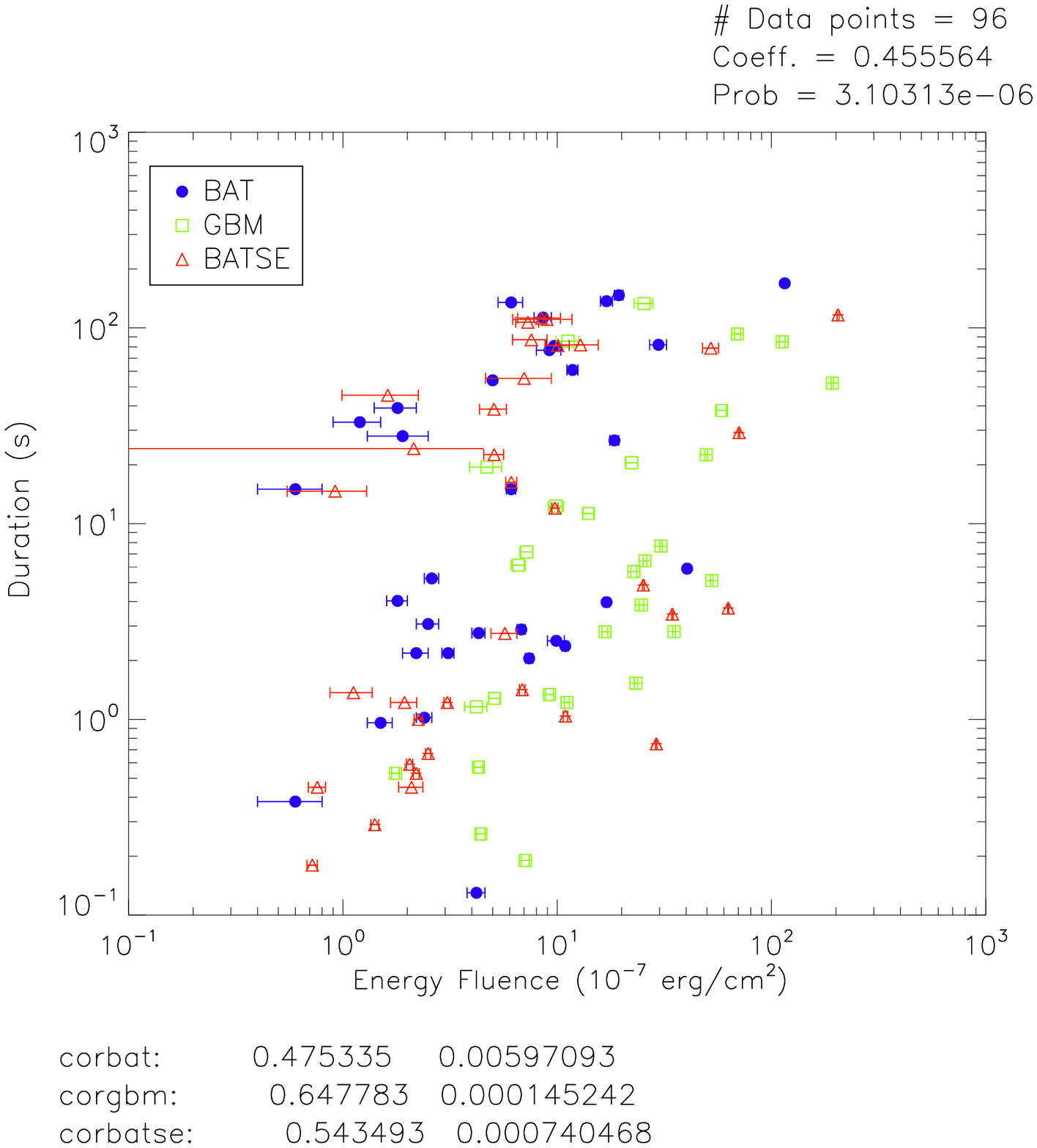} \caption{Energy fluence vs.~duration
for all of the
  events.  A correlation was found in all datasets individually.
  A correlation with increased significance ($P \sim 10^{-6}$) were found when
  all datasets were combined. } \label{dur_flnc}
\end{figure}

\begin{figure*}
  \centering
\includegraphics[trim=0cm 0cm 0cm 0.9cm, clip=true, scale=0.4]
{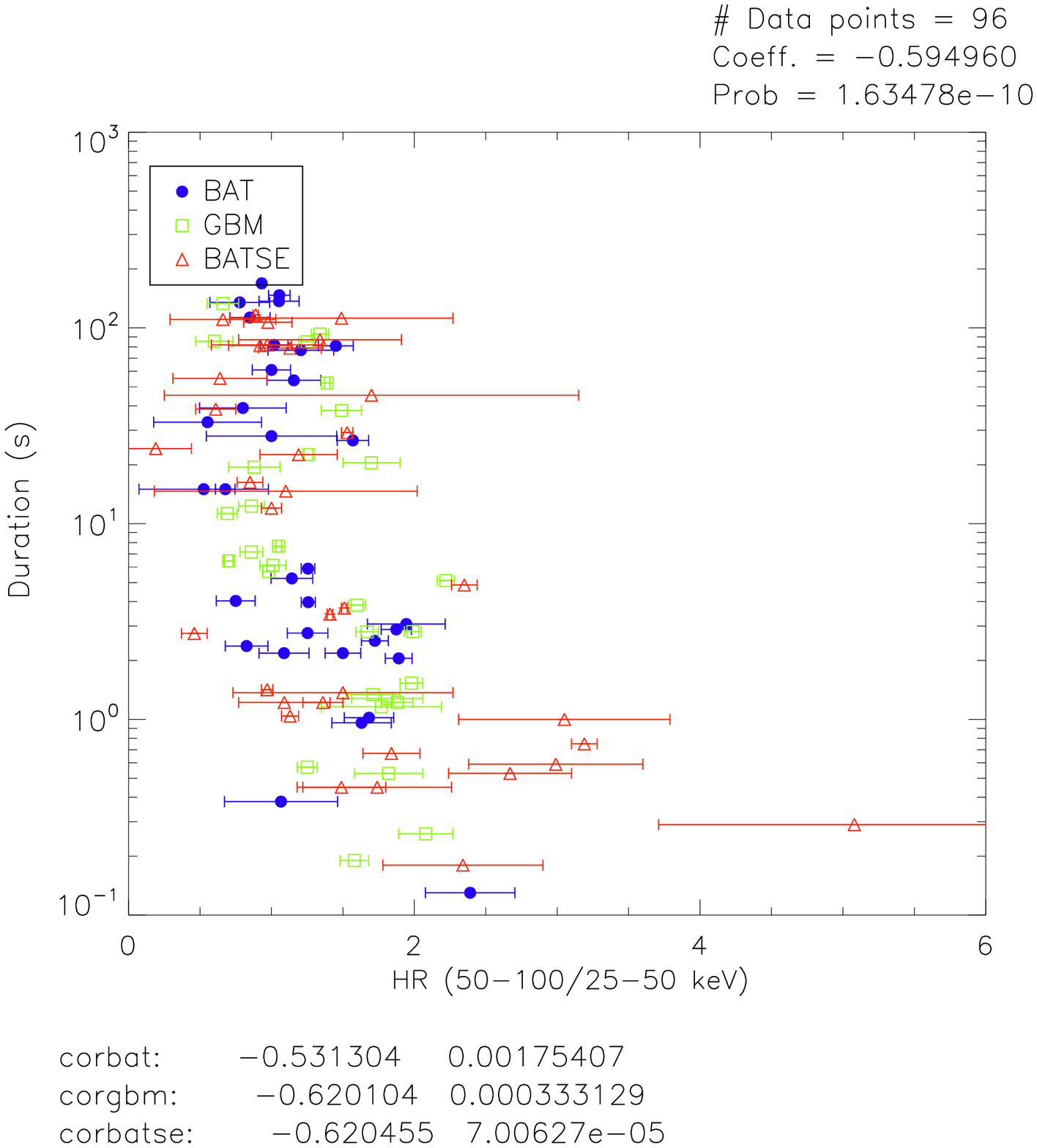}
\includegraphics[trim=0cm 0cm 0cm 0.9cm, clip=true, scale=0.4]
{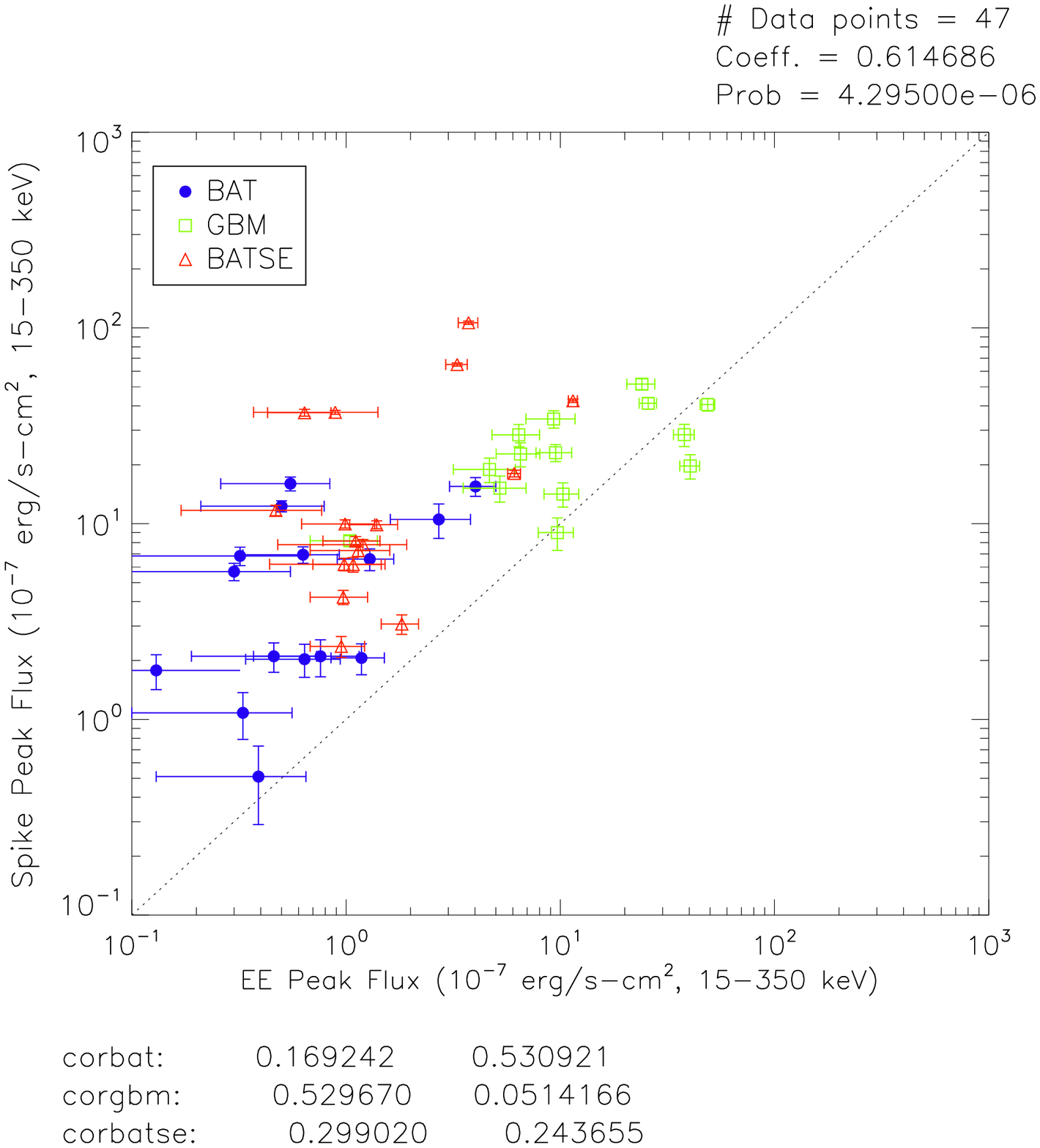}
\includegraphics[trim=0cm 0cm 0cm 0.9cm, clip=true, scale=0.4]
{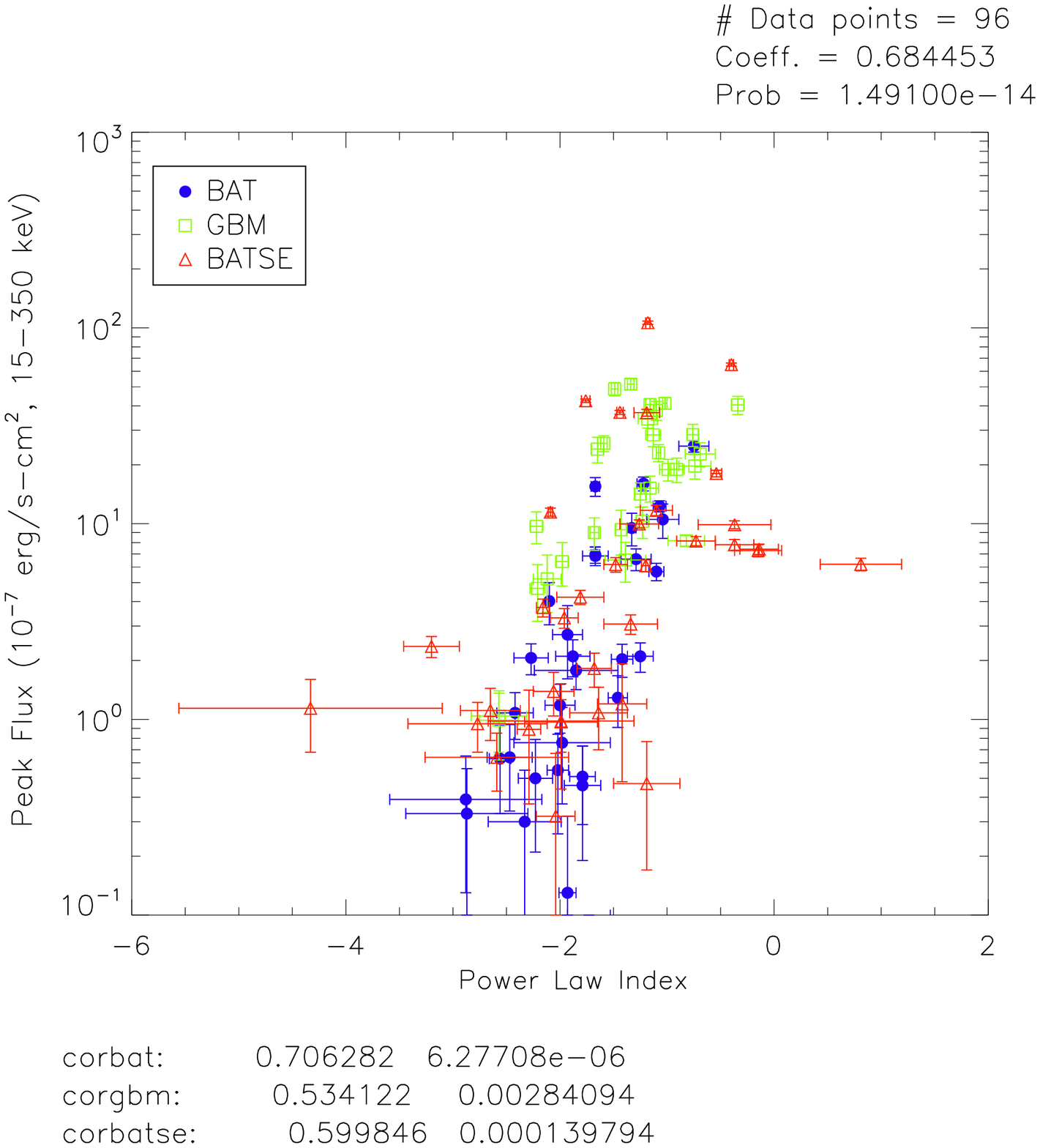}
\caption{[$Top$] Duration ($T_{\rm spike}$ and $T_{\rm EE}$)
vs.~hardness ratio ($left~panel$), spike peak flux vs.~EE peak flux
($right~panel$) for all GRBs with EE identified in BATSE, BAT, and
GBM data.  The dotted line shows the 1:1 line. [$Bottom$] Peak flux
and spectral indices for both spike and EE components of all events.
In all of the three parameter pairs, correlations with increased
significance ($P \sim 10^{-14} - 10^{-6}$) were found when all
datasets were combined.} \label{dur_hr}
\end{figure*}

We list in Table~\ref{corr_tab} only the pairs of parameters between
which a significant correlation (with a chance probability, $P
\lesssim 10^{-4}$) was found in at least one of the datasets of BAT,
GBM, or BATSE.  Only the correlation between the duration and energy
fluence is `universal', found to be significant in all three data sets 
(see Table~\ref{corr_tab} and Figure~\ref{dur_flnc}).
However, including this one, the significance of many correlations
greatly improved when all datasets are combined.
 The most significant combined correlations are found between peak flux and spectral index ($P \sim 10^{-14}$), duration and hardness ratio ($P \sim
10^{-10}$), and peak flux and hardness ratio ($P \sim 10^{-9}$).
 The correlations between peak flux of spike and EE ($P \sim
 10^{-6}$) also became significant only when all data were combined (see
Figure~\ref{dur_hr}).
It is possible that such an improvement is artificially introduced by putting together
samples, each of which has some degree of systematic biases in
observed parameters.   Uncertainties in cross
calibrations among these three instruments may also be a
contributing factor. Nonetheless, the most significant combined
correlations such as the ones mentioned above (also shown in Figure~\ref{dur_hr}) may indicate that the correlations exist
when including wider ranges of the parameters, and that each of
these instruments is detecting different population in these
parameter space. These correlations simply became statistically much
more significant with more data points. As we showed in the
simulations in \S\ref{sec:sim}, the EE detection sensitivities of
BAT and GBM differs by an order of magnitude, which may explain the
scatter in the peak flux values of among the three samples.
\begin{table*}
  \centering  \setlength{\tabcolsep}{0.04in} \caption{List of
    parameter pairs that are significantly correlated in at least one of
    the event samples of BAT, GBM, or BATSE.  ALL means BAT+GBM+BATSE
    all combined. Spearman's rank-order correlation coefficient, $\rho$
    with a chance probability, $P$, are shown.  $N$ is the number of
    data points.  The boldface indicates where significant correlations
    ($P < 10^{-4}$) were found.}
\label{corr_tab}
\begin{tabular}{cc|cccc|cccc|cccc|ccc}
\hline\hline

& & \multicolumn{3}{c|}{BAT} &&
\multicolumn{3}{c|}{GBM} && \multicolumn{3}{c}{BATSE}
&& \multicolumn{3}{c}{ALL} \\
\cline{3-5} \cline{7-9} \cline{11-13} \cline{15-17} \\

\multicolumn{2}{c|}{Parameters}& $\rho$ & $P$ & $N$ && $\rho$ & $P$&
$N$ && $\rho$ & $P$ & $N$ && $\rho$ & $P$ & $N$ \\
\hline
Duration & Spectral Lag$^\#$ & {\bf 0.68} & {\bf 1.4E$-$4} & 26 &&0.46 &2.0E$-$2 & 25 && 0.18&4.2E$-$1& 22 && {\bf 0.54 }&   {\bf 9.4E$-$7}  &  73 \\
Duration & Peak Flux & {\bf $-$0.56} & {\bf 9.5E$-$4} & 32  &&$-$0.27 & 1.5E$-$1  &  29 && {\bf$-$0.63} & {\bf 5.2E$-$5}& 35 && {\bf $-$0.52}&{\bf 5.8E$-$8}  &  96 \\
Peak Flux & Hardness Ratio$^*$ & {\bf 0.70} &  {\bf 7.3E$-$6} & 32 &&0.45  & 1.4E$-$2&  29 && 0.44 & 8.3E$-$3 & 35 &&{\bf 0.55} & {\bf 5.6E$-$9} &   96  \\
Duration & Hardness Ratio$^*$ & $-$0.46 & 7.6E$-$3 &   32 && $-$0.44&1.8E$-$2 &   29 && {\bf $-$0.56} & {\bf 4.6E$-$4} & 35 && {\bf$-$0.59} & {\bf 1.6E$-$10}  &  96 \\
Duration & Energy Fluence & 0.48 &  6.0E$-$3  &  32 && {\bf 0.65} &{\bf 1.4E$-$4} &   29 && {\bf 0.54} & {\bf 7.4E$-$4} & 35 && {\bf 0.46} &{\bf 3.1E$-$6} & 96  \\
Peak Flux & PWRL Index & {\bf 0.71} &  {\bf 6.3E$-$6} & 32 && 0.53& 2.8E$-$3&  29 && {\bf 0.60} & {\bf 1.4E$-$4} & 35 &&{\bf 0.68} & {\bf 1.5E$-$14} &   96  \\
Peak Flux, Spike & Peak Flux, EE & 0.17 & 5.3E$-$1 & 16  && 0.53 &5.1E$-$2 & 14 && 0.3 &   2.4E$-$1 & 17 && {\bf 0.61} &{\bf 4.3E$-$6} & 47 \\
Hardness Ratio$^*$, Spike & Hardness Ratio$^*$, EE&0.36&1.8E$-$1&16&&{\bf 0.82}&{\bf 2.9E$-$4}&14&&$-$0.01&9.7E$-$1&17&&0.27&6.4E$-$2&47\\

\hline
\end{tabular}
 \footnotesize{ \begin{flushleft}
$^\#$ Some spectral lag values are associated with large uncertainties,
which are not taken into account here (see Table~\ref{tab:lag}) \\
 $^*$ The hardness ratio are calculated in 50$-$100/25$-$50\,keV
for BAT, 100$-$300/50$-$100\,keV for GBM and BATSE events. For the
combined correlation (i.e., ALL), all ratio are calculated in
50$-$100/25$-$50\,keV.\\
\end{flushleft}}

\end{table*}

\begin{figure}
\centering
\includegraphics[trim=0cm 0cm 0cm 0.9cm, clip=true, scale=0.4]{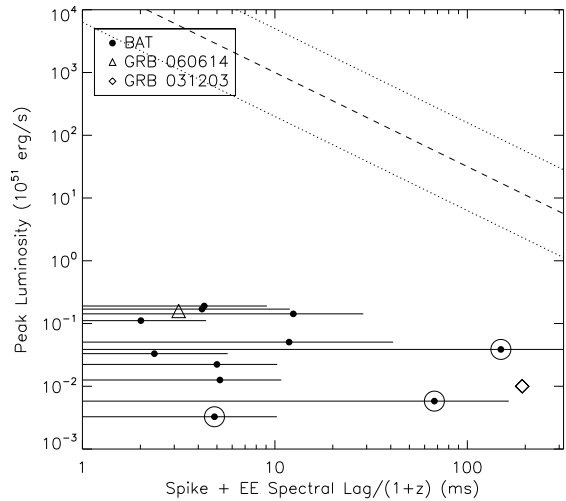}
\caption{Peak luminosity vs.~spectral lag for the 8 events with
  redshift values.  EE values available for 3 events are circled. The
  dashed line shows the correlation for long GRBs (a power law of
  index $-$1.5) with 1$\sigma$ confidence level shown as dotted
  lines \citep{ukw10}. The values of GRB\,060614 is also shown as a
  reference (triangle).}
\label{lag_lum}
\end{figure}

For the 8 events for which there are redshift measurements, we also
looked at source-frame parameters, such as redshift-corrected
durations, energy fluence, isotropic-equivalent total energy, and
luminosity.  All of these 8 events are best described with PWRL. The
sample size is small but we did not find any correlations using the
source-frame parameters. It is known that the peak luminosity and
the spectral lag are anti-correlated only for long GRBs
\citep{norris00, ukw10}.  We compare our sample's peak luminosity
and lags to the long-GRB correlation in Figure~\ref{lag_lum}. There
are 3 events among the 8 for which we were able to calculate
spectral lag of the EE, so those are also included in the plots.
These data points are circled. All of them, including the EE, lie
away from the correlation found for long GRBs but are consistent
with GRB\,060614 as well as other short GRBs.  A few of the EE with
longer lags are similar in values to sub-luminous long GRBs,
sometimes associated with supernovae; e.g., GRB\,031203.

\subsection{Spike vs. EE properties}
\begin{figure*}
\centering
\includegraphics[trim=0cm 0cm 0cm 0.9cm, clip=true, scale=0.4]
{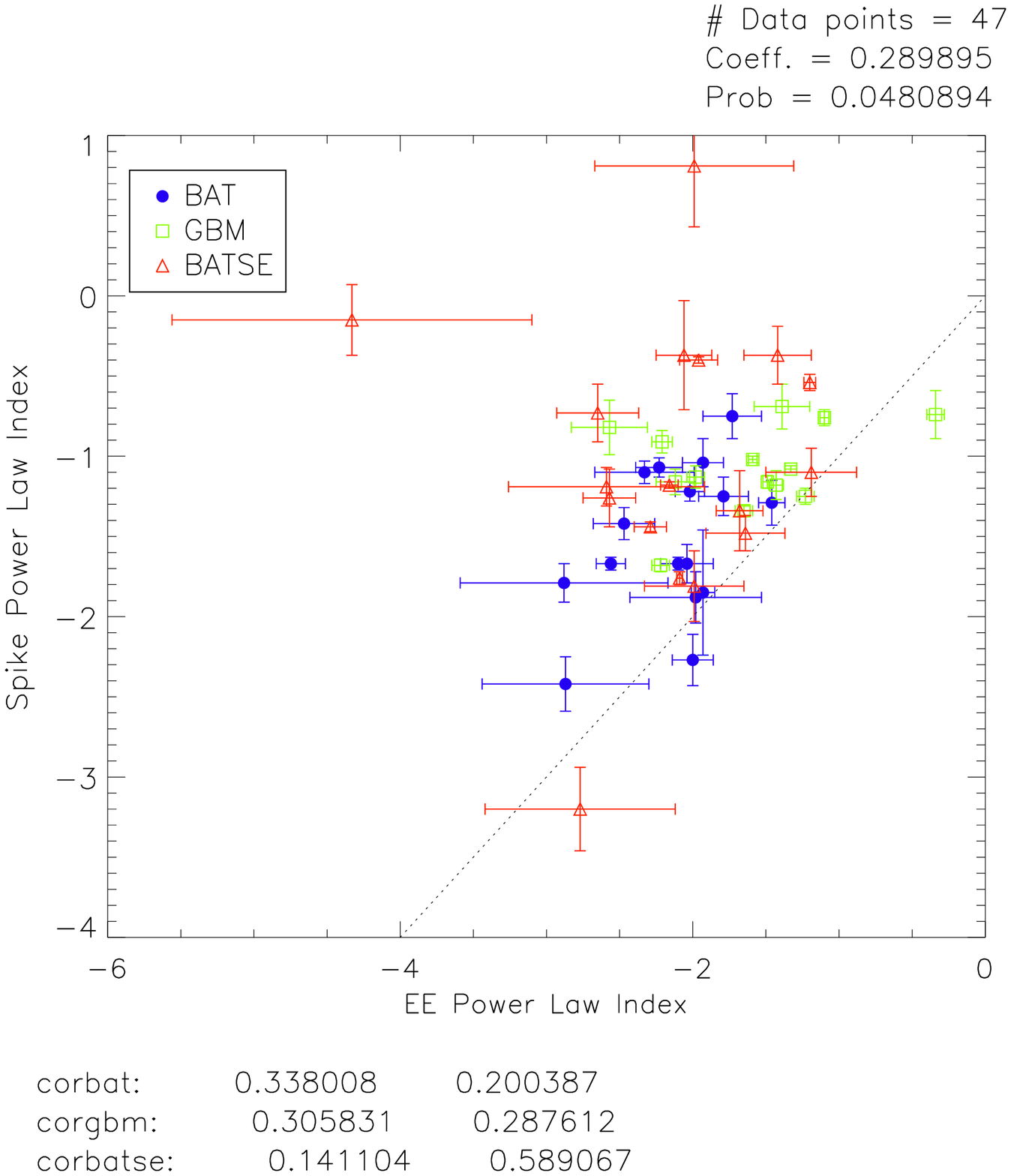}
\includegraphics[trim=0cm 0cm 0cm 0.9cm, clip=true, scale=0.4]
{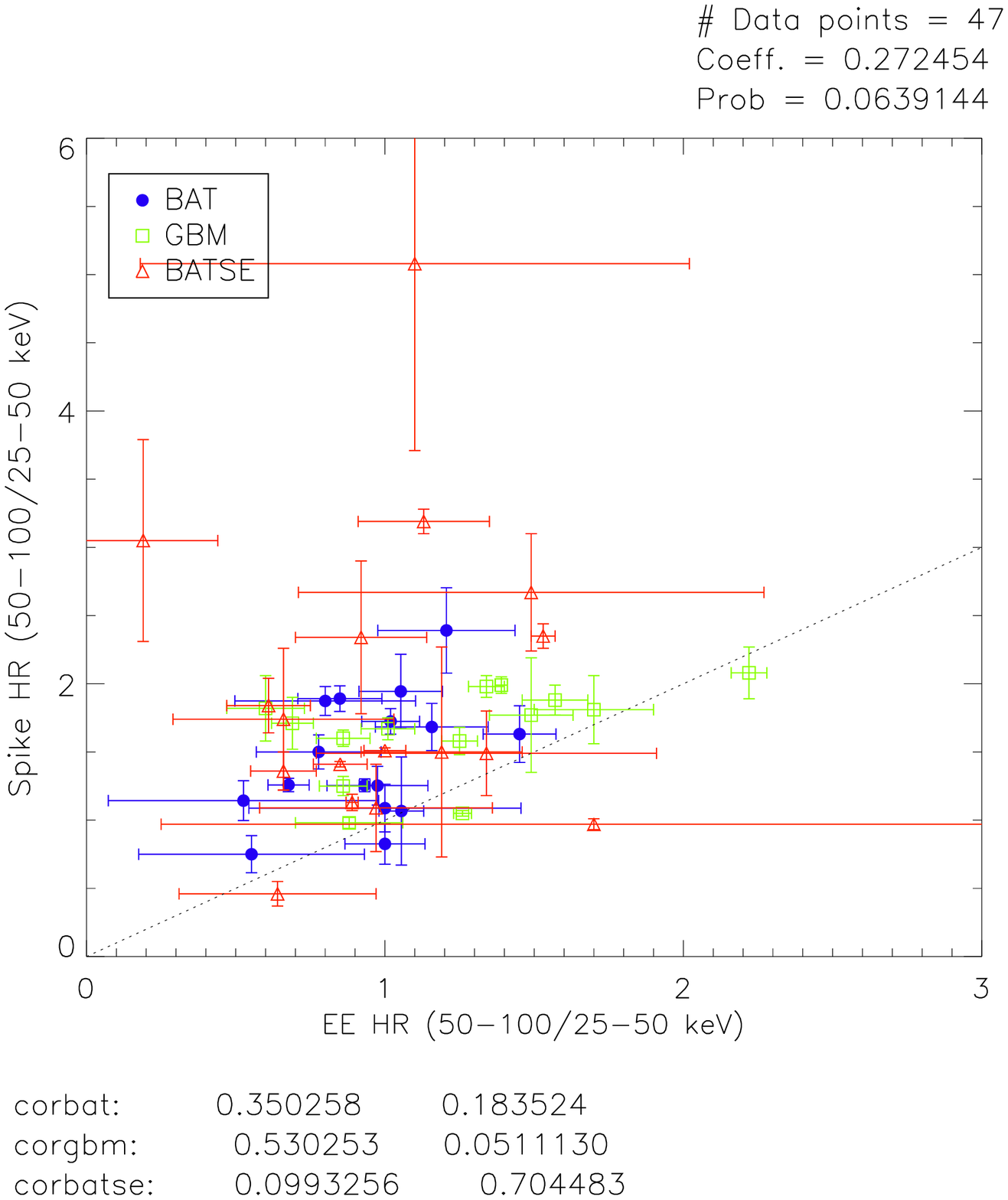} \caption{PWRL indices ($left$) and the
hardness ratio ($right$) of the
  spikes vs.~EE of the 30 candidate events. BAT and GBM events are
  shown with different symbols and colors. The 1:1 line is shown as dotted line.
  Although no correlations were found between these parameters, the plots indicate that the spikes tend to be spectrally harder than the EE.
  For comparison, the values
  for BATSE GRBs with EE from our previous work are also included.
    } \label{indsp_indee}
\end{figure*}

It has been reported in literature as well as in our previous study
of the BATSE GRBs with EE that the EE components tend to be
spectrally softer than the initial spikes.  Therefore, the spectral
indices of longer EE components are expected to be smaller than
those of the shorter spikes. The PWRL indices of the spikes and EE
(for all BAT, GBM, and BATSE samples) are compared in
Figure~\ref{indsp_indee}, in which we see a clear indication of EE
indices being softer, although no significant correlations were
found in any of the three samples.
We also compare in Figure~\ref{indsp_indee} the hardness ratio of
spike versus EE. We also see the indication of the spikes being
spectrally harder than the corresponding EE, but only within the GBM
sample was a significant positive correlation found (Spearman's
rank-order correlation coefficient, $\rho = 0.82$ with a chance
probability, $P \sim10^{-4}$, see Table~\ref{corr_tab}).

Furthermore, as stated earlier, the hardness ratio is negatively
correlated with duration, especially when all datasets are combined
($\rho = -0.59, P \sim 10^{-10}$, see also Figure~\ref{dur_hr}).
This reinforces the statement that the spikes are harder than the
EE, but additionally, this demonstrates that the anti-correlation
also exists within the spikes or EE components.

\section{Discussion}\label{sec:discussion}

We performed the most comprehensive investigations to date to search
for EE that follow short GRBs observed with {\it Swift} BAT and {\it
Fermi} GBM detectors.  Based on the data availability and the
filtering with the morphological criteria, the burst sample size
subjected to our systematic search for the EE was 128 BAT and 287
GBM GRBs. These correspond to 19\% and 28\% of the total number of
GRBs detected with BAT and GBM respectively, up until 2013 January
31. As a comparison, the fraction of GRBs classified as
short-duration ($<$ 2\,s) event within the entire GRB data sample is
13\% for BAT and 17\% for GBM \citep{sak11, vk14}. Our search
identified EE in 16 bursts seen with BAT, and 14 bursts with GBM. Of
these three events are common in both EE samples, and two more events in
the BAT samples were also observed with GBM but without the
EE identified in the search. We see at least one event
(GRB\,100212A) of which the EE seen in the BAT coincided with what
appeared as a X-ray flare in the XRT data.

None of these 30 candidates are considered ``short" GRBs, by the
conventional definition of $T_{90} < 2$\,s. Even the duration of the
spikes that we determined for our candidate events is mostly longer than
2\,s, although the duration was calculated here using the lower
energy ranges ($<$50\,keV) than the standard energy range in which
the duration is calculated for these instruments. Nonetheless, the
negligible spectral lag of the spikes we found is consistent with
short GRBs' properties. On the other hand, the duration of the EE
components span a range of $\sim$10\,s to $\sim$150\,s, and the
spectral lag we found for the EE is longer on average, being more
consistent with long GRBs'.

The ratio of the energy emitted in the EE and in the spike (i.e.,
energy fluence ratio, $E_{\rm EE}/E_{\rm Spike}$) ranges from 0.21
to 33.5 (see Table~\ref{tab_ratio}), with a median value of 1.7
(1.99 for BAT sample and 1.67 for GBM sample). This means that for
the majority of these events, more energy is emitted as EE than
during the initial burst spike, significantly more in some cases.
Interestingly, we found that $E_{\rm EE}$ is positively correlated
with $E_{\rm Spike}$, which was revealed only when all datasets are
considered together ($P = 7 \times 10^{-3}$).

We have also shown the cases where the burst is observed with both
BAT and GBM  (i.e., common event) but the EE was identified with
only one of the detectors due to the difference in their
sensitivities in a given energy range. Including also the cases
where the EE was identified in both detectors' datasets, the joint
spectral analysis yielded better constrained parameters.

\subsection{Comparison with the BATSE events}

Previously we identified 19 BATSE GRBs with EE
using the same systematic search algorithm that we applied to the
BAT and GBM data here \citep{bos13}.  The size of the sample subjected to the search was 296, i.e., 14\% of all GRBs detected with BATSE in its entire mission.
The number is actually smaller than the total number of short-duration BATSE GRBs (25\% of all GRBs), due to the lack of orbital data used for the background estimation.
This was not the case for the GBM data, and that increased the fraction of short-duration GBM GRBs included in the search sample.

Comparing the BATSE EE candidate events to the ones found here with
BAT and GBM using the same search method, could reveal (or confirm
the lack there of) any systematic differences between the
instruments or biased preference of a detector towards detecting
certain types of bursts.
In Figure~\ref{dur_hr} we have shown the plots of three parameter
pairs between which we found significant correlations.  When each
dataset (i.e., BAT, GBM, or BATSE) was individually studied, the
correlations are only significant ($P > 10^{-4}$) in one or two of
the datasets. Nevertheless, the correlations were found with much
higher significance when all datasets were combined
(Table~\ref{corr_tab}). There in the peak flux plots, we see that
most of the GBM events have higher peak flux in both spikes and EE
than those of BAT and BATSE. In the top right-panel plot where the
spike peak flux vs.~EE peak flux is shown, it is also clear that the
GBM events are with higher flux for both the spikes and the EE. This
was also noticeable in terms of energy fluence of the spikes and the
EE,  with the GBM events offering the higher energy fluence on average than the BAT and BATSE events (see Figure~\ref{dur_flnc}).

The comparison between the general population of GRBs observed with
GBM and with BATSE do not show much difference in their flux values
as well as their spectral characteristics \citep{nava11}. Therefore,
we suspect that the clear difference in the peak flux values found
between the BATSE and GBM GRBs with EE is due to the significant
difference in the detectors' effective area in the energy range
where the EE components are most prominent. BATSE Large Area
Detectors (as well as the BAT) had $\sim$10 times larger effective
area than that of GBM NaI detectors below $\sim$100\,keV
\citep{tie13, sta09} where the EE components were usually detected.
Thus, only the bursts associated with bright (i.e., higher peak
flux) EE components were identified in our search using the GBM
data. Then, the fact that the spike peak flux of these GBM events
are also higher than the other (BAT and BATSE) events' is probably
due to the strong positive correlation between the peak flux of the
spikes and that of the EE components.

\subsection{Two-Component Jet Model}
In our previous study with the BATSE sample, we examined the
two-component jet model of EE proposed by \citet{bar11}, by estimating
the ratio of the opening angles of the two jet components
($\theta_{BZ}/\theta_{\nu\tilde{\nu}}$) for each of the candidate GRBs
with EE.  The angles were estimated based on the relation:
\begin{equation}
{{L_{BZ}}\over{L_{\nu\tilde{\nu}}}}~\left(\frac{\theta_{\nu\tilde{\nu}}}
{\theta_{BZ}} \right)^2 {{t_{BZ}}\over{t_{\nu\tilde{\nu}}}}=
{{E_{\rm EE}}\over{E_{\rm spike}}},
\end{equation}
where $E_{\rm EE}$ and $E_{\rm spike}$ are the observed energy
fluence, and $t_{\nu\tilde{\nu}}$ and $t_{BZ}$ are durations of the
spike and the EE, respectively. Following \citet{bar11}, we used the
luminosity estimates of $L_{BZ} \approx 10^{48}$\,erg\,s$^{-1}$ and
$L_{\nu\tilde{\nu}} \approx 3\times10^{50}$\,erg\,s$^{-1}$ derived
assuming typical physical parameters of the progenitor.

\begin{table}
  \centering \caption{Energy fluence ratio and the corresponding opening angle ratio estimates of the two
    jet components. GRB\,111228 is excluded due to the insufficient statistics of its EE component
    to determine the energy fluence.} \label{tab_ratio}
\begin{tabular}{lcc}
\hline GRB name & $E_{\rm EE}/E_{\rm Spike}$ &$\theta_{BZ}$/$\theta_{\nu\tilde{\nu}}$ \\
        & (15--350keV) & \\
\hline
{\bf BAT} &     &         \\
050724  &   1.69    &   0.28    \\
051016B &   0.64    &   0.21    \\
060614  &   4.43    &   0.15    \\
061006  &   1.16    &   0.40 \\
061210  &   2.20 &   0.95    \\
070506  &   0.22    &   0.21    \\
070714B &   0.26    &   0.41    \\
080503  &   33.5   &   0.20 \\
090531B &   2.08    &   0.29    \\
090927  &   0.87    &   0.22    \\
100212A &   1.99    &   0.32    \\
100522A &   0.36    &   0.19    \\
110207A &   6.69    &   0.15    \\
110402A &   3.01    &   0.19    \\
111121A &   1.09    &   0.28    \\
121014A &   6.28    &   0.21    \\
{\bf GBM} &     &         \\
080807993   &   1.66    &   0.21    \\
090131090   &   1.63    &   0.08    \\
090820509   &   1.67    &   0.16    \\
090831317   &   15.8   &   0.31    \\
091120191   &   5.47    &   0.11    \\
100517154   &   1.53    &   0.14    \\
100522157   &   0.40 &   0.16    \\
110207470   &   4.41    &   0.15    \\
110402009   &   14.0   &   0.09    \\
110824009   &   2.97    &   0.26    \\
120402669   &   0.21    &   0.23    \\
120605453   &   0.39    &   0.14    \\
121029350   &   12.1   &   0.07    \\

\hline
\end{tabular}
\label{modelresults}
\end{table}


For comparison, we also estimate the jet opening angle ratios for
our sample here, which we present in Table~\ref{tab_ratio}. We found
that the angle ratio estimates with the BAT and GBM samples were
smaller than those of the BATSE sample.
This is likely due to the fact that the fluence ratio
($E_{\rm EE}$/$E_{\rm spike}$) of BAT and GBM samples are
larger on average than those of the BATSE sample.
Moreover, the BATSE GRBs with EE
  have shorter $T_{\rm spike}$ and longer $T_{\rm EE}$ on average
  than the BAT and GBM GRBs with EE, which makes
  $t_{BZ}$/$t_{\nu\tilde{\nu}}$ smaller for BAT+GBM samples and thus,
  $\theta_{BZ}$/$\theta_{\nu\tilde{\nu}}$ also smaller.

The angle ratio of the BAT and GBM GRBs estimated here span from
 0.07 to 0.95, with a median value of 0.21.
(they were 0.05 to 0.67, and 0.29 median for the BATSE sample). The
neutorino-heated ($\nu\tilde{\nu}$) jet is expected to have an
opening angle of $\sim$0.1, and the electromagnetic Blandford-Znajek
(BZ) jet should have an opening angle that is inversely proportional
to the Lorentz factor; $\theta_{BZ} \sim 1/\Gamma_{BZ}$.
 Then, the ratios we found here correspond to
$\Gamma_{BZ}$ of the order of 10$-$100, which is still in the range
expected for the Lorenz factor of evolving BZ jet \citep{bar11}.

\section*{Acknowledgments}
We thank the anonymous referee for his/her insightful and
constructive suggestions, which significantly improved the paper.
This project was supported by the Scientific and Technological
Research Council of Turkey (T\"UB\.ITAK grant 109T755).

\end{document}